\documentclass[aps,prd,twocolumn,nofootinbib,floatfix]{revtex4-1}
\usepackage{setspace}
\usepackage{amsthm}
\theoremstyle{definition}

\usepackage{graphicx}
\usepackage{dcolumn}
\usepackage{multirow}
\usepackage{subfigure}
\usepackage{times,mathptm}
\usepackage{float}
\usepackage{color}
\usepackage{amsmath,amsfonts}
\usepackage{mathptmx}
\usepackage{mathrsfs}
\usepackage{bbm}
\usepackage{bm}
\usepackage{xfrac}

\newcommand{\beq}{\begin{equation}}
\newcommand{\eeq}{\end{equation}} 
\newcommand{\bea}{\begin{eqnarray}}
\newcommand{\eea}{\end{eqnarray}}

\newcommand{\ua}{\uparrow}
\newcommand{\da}{\downarrow}
\newcommand{\teta}{\widetilde{\eta}}

\newcommand{\Sc}{S$_\text{c}$}
\newcommand{\Hs}{H_\text{spin}}
\newcommand{\Zs}{Z_\text{spin}}

\renewcommand{\d}{\delta}
\newcommand{\ihat}{\boldsymbol{\hat{\textbf{\i}}}}
\renewcommand{\l}{\lambda}

\newcommand{\ophi}{\overline{\phi}}

\newcommand{\hF}{\widehat{F}}

\renewcommand{\b}{\beta}
\renewcommand{\a}{\alpha}

\renewcommand{\ni}{\noindent}
\newcommand{\tr}{\text{Tr}}

\newcommand{\vx}{{\vec{x}}}
\newcommand{\vy}{{\vec{y}}}
\newcommand{\vz}{\vec{z}}

\newcommand{\n}{\nu}
\newcommand{\m}{\mu}
\newcommand{\q}{\overline{q}}
\newcommand{\g}{\gamma}

\newcommand{\e}{\epsilon}
\newcommand{\s}{\sigma}
\renewcommand{\k}{\kappa}
\newcommand{\D}{\Delta}

\newcommand{\N}{{\cal N}}

\renewcommand{\th}{\theta}

\newcommand{\oh}{\frac{1}{2}}

\newcommand{\dg}{\dagger}
\newcommand{\non}{\nonumber}

\newcommand{\rf}[1]{(\ref{#1})}
\newcommand{\ra}{\rightarrow}
\newcommand{\pa}{\partial}
\renewcommand{\vec}[1]{\bm #1}

\usepackage{ulem}

\begin{document}

\title{The Higgs phase as a spin glass, and the transition between varieties of confinement} 

\bigskip
\bigskip

\author{Jeff Greensite and Kazue Matsuyama}
\affiliation{Physics and Astronomy Department \\ San Francisco State
University  \\ San Francisco, CA~94132, USA}
\bigskip
\date{\today}
\vspace{60pt}
\begin{abstract}

\singlespacing
 
    We propose that the Higgs phase of a gauge Higgs theory is the phase of spontaneously broken
custodial symmetry, and present a new gauge invariant order parameter for custodial symmetry breaking which is very closely analogous to the Edwards-Anderson order parameter for spin glasses.  Custodial symmetry is a global symmetry
acting on the Higgs field alone, and we show here that the spin glass transition in gauge Higgs theories, from a QCD-like phase to a Higgs phase of broken custodial symmetry, coincides with the transition between two distinct types of confinement.  These are color confinement in the Higgs phase, and a stronger version of confinement, which we have termed ``separation-of-charge'' confinement, in the QCD-like phase.

\end{abstract}

\maketitle
 
\singlespacing

\section{\label{Intro} Introduction}

   In a gauge Higgs theory, with the scalar field in the fundamental representation of the gauge group, there
is no thermodynamic transition which entirely separates the phase diagram into a confining and a Higgs phase 
\cite{Osterwalder:1977pc,Fradkin:1978dv}.   There
is nonetheless a confining region which is qualitatively much like QCD, in the sense that there are metastable
color electric flux tubes which break by pair creation, and, as a consequence, linear Regge trajectories whose resonances correspond to such metastable states.  There is also a Higgs region, in which all forces mediated by bosons are 
Yukawa in character, as in the weak interaction sector of the Standard Model, and there is no flux tube formation at any length scale.  The physics in these two regions, like the physics of QCD and the weak interactions, is qualitatively so different that one may ask whether the distinction may be formulated precisely, despite the absence of a thermodynamic separation.   And, if there is such a physical distinction, one may also ask whether it is associated with the symmetric or broken realization of some symmetry.   

   In previous work \cite{Greensite:2017ajx} we have suggested that the Higgs and confining regions are distinguished by different varieties of confinement, namely color (C) confinement in the Higgs region and a stronger type of confinement, which we call ``separation of charge'' (\Sc) confinement, in the QCD-like region, and we have shown that there must be a sharp transition between these types of confinement \cite{Greensite:2018mhh}.  We have also suggested
that this transition might coincide with custodial symmetry breaking, and put forward a gauge invariant criterion for
such breaking.  Custodial symmetry is a group whose elements transform the Higgs field but not the gauge field, and
for a Higgs in the fundamental color representation this group contains, at a minimum, the center elements of the gauge group.  As an example, we consider the lattice SU(2) gauge Higgs theory with a unimodular Higgs field in the fundamental representation of the gauge group.  The action is
\bea
     S &=& S_W[U] + S_H[\phi,U]  \non \\
        &=& - \beta \sum_{plaq} \oh \mbox{Tr}[U_\m(x)U_\n(x+\hat{\m})U_\m^\dg(x+\hat{\n}) U^\dg_\n(x)]  \non \\
   & &       - \gamma \sum_{x,\m} \oh \mbox{Tr}[\phi^\dg(x) U_\m(x) \phi(x+\widehat{\m})] \ ,
\label{Sgh}
\eea
where $\phi$ is an SU(2) group-valued field.  This theory has the following invariances:
\bea
            U_\m(x) &\ra& L(x)  U_\m(x) L^\dg(x+\hat{\mu})  \non \\
            \phi(x) &\ra& L(x) \phi(x) R \ ,
\label{LR}
\eea
where $L(x) \in $ SU(2)${}_{gauge}$ is a local gauge transformation, while $R \in $ SU(2)${}_{global}$ is a global transformation.  SU(2)${}_{global}$ is the custodial symmetry group.\footnote{The term ``custodial symmetry'' is drawn from the electroweak theory \cite{Willenbrock:2004hu,Weinberg:1996kr},  and has been applied to other beyond-the-standard-model theories, see e.g.\ \cite{Maas:2019nso}.  The term is actually defined in different ways (see section \ref{diagonal}), but for our purposes (following \cite{Georgi:1994qn,Maas:2019nso})  it refers, in a gauge-Higgs theory, to a group of transformations of the scalar field alone which leaves the action invariant. } Likewise, in the abelian-Higgs model or the SU(3) gauge-Higgs model, where $\phi$ has one or three color components respectively,  the custodial group consists of the 
global U(1) transformations ${\phi(x) = e^{i\th} \phi(x)}$.   

    The purpose of this article is to introduce a new criterion for custodial symmetry breaking which is very closely analogous to the Edwards-Anderson order parameter for spin glass transitions \cite{Edward_Anderson}, and to show that custodial symmetry breaking according to the new criterion coincides physically with the transition from \Sc \ to C confinement.  This implies that the Higgs regime of a gauge Higgs theory can be regarded as a spin glass phase.

\section{\label{spin}The spin glass order parameter}

    Just as $\langle \phi \rangle$ vanishes in a gauge Higgs theory in the absence of gauge fixing, so the expectation
value of Ising spins vanishes in a spin glass.  The reasons are similar.  The Edwards-Anderson spin glass model
\cite{Edward_Anderson} is described by the Hamiltonian
\beq
           \Hs =  - \sum_{ij} J_{ij} s_i s_j - h \sum_i s_i \ ,
\label{EA}
\eeq
where $s_i=\pm 1$ is an Ising spin at site $i$, $J_{ij}$ are random couplings between spins at sites $i,j$ (which may or may not be nearest neighbors) drawn from some probability distribution $P(J)$, and $h$ represents an external magnetic field.  At $h=0$ the model is obviously symmetric
under the global $Z_2$ symmetry $s_i \ra z s_i, ~ z=\pm 1$.  But because of the random nature of the couplings $J_{ij}$,
the spatial average of spins vanishes in the $h=0$ limit, as does the expectation value of any individual spin, upon averaging over the random couplings.  Despite this fact there is a way to detect the spontaneous breaking of the global $Z_2$ symmetry.  Define
\bea
           \Zs(J) &=&  \sum_{\{s\}} e^{-\Hs/kT}  \label{sg1} \\
           \overline{s}_i(J) &=& {1\over \Zs(J)}  \sum_{\{s\}} s_i e^{-\Hs/kT} \\ 
           q(J) &=& {1\over V} \sum_i  \overline{s}^2_i(J)   \\
           \langle q \rangle &=& \int \prod_{ij} dJ_{ij} \ q(J) P(J) \label{sg4} \ ,
\eea           
where $q(J)$ is the Edwards-Anderson order parameter.  
When the expectation value $\langle q \rangle$ is non-zero in
the infinite volume $V \ra \infty$ and $h \ra 0$ limits, the system is in the spin glass phase, and the $Z_2$ global symmetry
is spontaneously broken.  Note that because $q(J)$ is a sum of squares it is actually invariant under
$Z_2$ transformations at $h \ra 0$.  Nevertheless, $q(J)$ detects whether the spins $s_i$, which do transform under this symmetry, will tend to have a particular orientation at each site $i$ at fixed couplings $J_{ij}$.  While the spatial average of spins will in general vanish in the infinite volume limit at $h\ra 0$, the average of a spin at any given site might not, and this is the symmetry breaking which is detected by a non-zero $q(J)$.

   In the analogous construction in lattice gauge Higgs theory, $\phi(\vx)$ has the role of the spin variables, with link variables 
$U_i(\vx)$ as the random couplings.  As in the case of the spin glass, the spatial average of $\phi(\vx)$ averages to zero
on a large volume, in the absence of gauge fixing, in any typical configuration.  Also as in a spin glass, the scalar field at any particular point $\vx$ averages to zero in a large set of  sample configurations $\{\phi,U_i\}$.  However, once again as in the spin glass situation, there is a meaningful sense in which a certain global symmetry, in this case custodial symmetry, 
can be said to have broken spontaneously.

We will continue to use SU(2) gauge Higgs theory as an illustration, and this is the theory which we numerically simulate, but the reasoning can be readily extended to the U(1) and other  SU(N) gauge groups.  The unimodular restriction $|\phi|=1$ is only a convenience which allows us to plot phase diagrams in a two dimensional $\b-\g$ plane. For the present we will restrict our considerations to simple U(1) and SU(N) gauge groups, with a single Higgs field in the fundamental (or, for U(1), the single charged) representation of the gauge group.

Let $H$ be the Hamiltonian operator of gauge Higgs theory in temporal gauge, a gauge chosen so that all physical states are gauge invariant.  We begin from
\bea
           \exp[-H(\phi,U)/kT] &=& \langle \phi,U| e^{-H/kT} |\phi,U\rangle \non \\
                                           &=& \sum_n |\Psi_n(\phi,U)|^2 e^{-E_n/kT}  \ ,
\label{H}
\eea
where the $\Psi_n$ are energy eigenstates.  It is straightforward, from a Euclidean path integral representation,
to derive the invariance of $H(\phi,U)$ under gauge and custodial symmetry transformations from the
corresponding symmetries of the action, see \rf{Hcust} below.  In analogy to spin models, we insert a small custodial symmetry breaking term
\beq
        H_{spin}(\phi,U,\eta) = H(\phi,U) -  h \sum_{\vx} \tr[\eta^\dg(\vx) \phi(\vx)]
\label{Hspin}
\eeq
with $\eta(\vx)$ an SU(2)-valued field.  We then define
\bea                        
            \Zs(U,\eta) &=& \int D\phi(\vx) \ e^{-\Hs(\phi,U,\eta)/kT}   \label{gg1} \\                     
           \overline{\phi}(\vx;U,\eta) &=& {1 \over  \Zs(U,\eta) }  \int D\phi \ \phi(\vx)  e^{-\Hs(\phi,U,\eta)/kT}  \label{overline} \\ 
            \Phi(U) &=& {1\over V} \left[\sum_\vx  | \overline{\phi}(\vx;U,\eta | \right]_{\eta \in \N(U)} \label{max}  \\
            \langle \Phi \rangle &=& \int DU_i(\vx) \ \Phi(U) P(U) \ ,
\label{gglass}
\eea
which should be compared to eqs.\ [\ref{sg1}-\ref{sg4}].  Here $|\overline{\phi}(x)|$ denotes the gauge invariant modulus, e.g.
\bea
|\overline{\phi}(\vx)|&=&\sqrt{\oh \tr \overline{\phi}^\dagger(\vx)\overline{\phi}(\vx)}  ~~~ \text{SU(2) gauge-Higgs theory}  \non \\
|\overline{\phi}(\vx)|&=&\sqrt{\overline{\phi}^\dg(\vx) \cdot \overline{\phi}(\vx)}~~~~~\text{SU(N) gauge Higgs theory} \ , \non \\
\eea
and $P(U)$ is a gauge invariant probability distribution for the link variables, described below, which is obtained from the
partition function after integrating out the scalar field.  The expression $\N(U)$ represents a set $\eta(\vx)$ fields defined by
\beq
     \N(U) = \underset{\eta}{\arg\max} \sum_\vx \left| \int D\phi \ \phi(\vx)  e^{-\Hs(\phi,U,\eta)/kT} \right| \ ,
\label{argmax}
\eeq
and the elements of this set are related by custodial transformations, as shown below.

   We now prove that the order parameter $\Phi(U)$ is gauge invariant, and independent of the choice of $\eta$ in the set $\N(U)$.  We begin by showing that $\Zs(U,\eta)$ is invariant under ${\eta \ra \eta R}$, where $R$ is an element of the custodial symmetry group.  Denote $\phi'(\vx) = \phi(\vx)R$,
and using the invariance of the measure and $H$ under custodial transformations,
\bea
   & &  \Zs(U,\eta R) \non \\
   & & ~ = \int D\phi' \exp[-(H(\phi',U) - h\sum_\vx \tr[R^\dg(\vx) \eta^\dg(\vx) \phi'(\vx)])/kT ] \non \\
   & & ~ = \int D\phi \exp[-(H(\phi,U) - h \sum_\vx \tr[\eta^\dg(\vx) \phi(\vx)])/kT ] \non \\
   & & ~ =  \Zs(U,\eta) \ .
\label{ZsR}
\eea
Likewise, denoting
\beq
\eta'(\vx) \equiv g(\vx) \eta(\vx) ~~,~~\phi'(\vx) \equiv g(\vx) \phi(\vx) \ ,
\label{chgvar}
\eeq 
where $g(\vx)$ is a local gauge transformation, we have
\bea
    & & \Zs(g\circ U,g\circ \eta) \non \\
    & & ~ =  \int D\phi' e^{-(H[\phi',g\circ U] - h\sum_\vx \tr[\eta'^\dg(\vx) \phi'(\vx)])/kT} \label{Zg1} \\
    & & ~ =   \int D\phi' e^{-(H[g\circ\phi,g\circ U] - h\sum_\vx \tr[\eta'^\dg(\vx) g(\vx)\phi(\vx)])/kT} \label{Zg2} \\
    & & ~ =  \int D\phi e^{-(H[\phi,U] - h\sum_\vx \tr[\eta^\dg(\vx) \phi(\vx)])/kT} \label{Zg3} \\
    & & ~ = \Zs(U,\eta) \ .
\label{Zsg}
\eea
By the same reasoning, we see that $\ophi$ transforms covariantly under transformations in the gauge and custodial symmetry groups.  Again denoting $\phi'(\vx) = \phi(\vx)R$,
\bea
 & &  \ophi(\vx;U,\eta R)  \non \\
 & & ~ = {1 \over \Zs(U,\eta R)} \int D\phi' \ \phi'(\vx)  e^{-(H[\phi',U] - h\sum_\vx \tr[R^\dg \eta^\dg(\vx)  \phi'(\vx)])/kT} \non \\
 & & ~ = {1 \over \Zs(U,\eta)} \int D\phi \ \phi(\vx) R \  e^{-(H[\phi,U] - h\sum_\vx \tr[\eta^\dg(\vx) \phi(\vx)])/kT} \non \\
 & & ~ =   \ophi(\vx;U,\eta) R \ ,
\label{ophiR}
\eea
which establishes covariance under custodial symmetry.  Again applying the change of variables \rf{chgvar}
we have also
\bea
 & &  \ophi(\vx;g\circ U,g\circ \eta) \non \\
 & & \qquad = {1 \over \Zs(g\circ U,g\circ \eta)} \int D\phi'~ \phi'(\vx)  e^{-H[\phi',g\circ U]/kT}  \non \\  
 & & \qquad \qquad \qquad \times    {e^{h\sum_\vx \tr[\eta^\dg(\vx) g^\dg(x) \phi'(\vx)])/kT}} \non \\
 & & \qquad = {1 \over \Zs(U,\eta)} \int D\phi ~ g(\vx)\phi(\vx)  e^{-H[\phi,U]/kT} \non \\
 & & \qquad \qquad \qquad \times    {e^{ h\sum_\vx \tr[\eta^\dg(\vx) \phi(\vx)])/kT}} \non \\
 & & \qquad = g(\vx) {1 \over \Zs(U,\eta)}  \int D\phi ~ \phi(\vx)  e^{-(H[\phi,U]/kT} \non \\
 & & \qquad \qquad \qquad \times  {e^{ h\sum_\vx \tr[\eta^\dg(\vx) \phi(\vx)])/kT}}    \non \\
 & & \qquad =  g(\vx) \ophi(\vx;U,\eta) \ .
 \label{ophig}
\eea

   The same changes of variables show that
\bea
& & \int D\phi \ \phi(\vx)  e^{-\Hs(\phi,U,\eta R)/kT} \non \\
& & \qquad = \left\{\int D\phi \ \phi(\vx)  e^{-\Hs(\phi,U,\eta)/kT} \right\} R \ ,
\eea
and
\bea
& &\int D\phi \ \phi(\vx)  e^{-\Hs(\phi,g\circ U,g \circ \eta)/kT} \non \\
& & \qquad = g(\vx) \left\{ \int D\phi \ \phi(\vx)  e^{-\Hs(\phi,U,\eta)/kT} \right\} \ .
\label{equiv}
\eea
These two relations, applied to \rf{argmax}, imply that:  \\

\ni  If ${\eta(\vx) \in \N(U)}$ then 
\begin{enumerate}
\item ${\eta(\vx) R \in \N(U)}$,
\item ${g(\vx) \eta(\vx) \in \N(g\circ U)}$. 
\end{enumerate}

\ni From point 1, and from \rf{ophiR}, we see that $\Phi(U)$ is independent of the choice of $\eta$ in the set $\N(U)$, since
these elements are related by transformations in the custodial group.  Then
it follows from point 2, and from \rf{ophig}, that
\bea
 \Phi(g\circ U) &=& {1\over V} \left[ \sum_\vx   |\overline{\phi}(\vx;g\circ U,\eta')| \right]_{\eta' \in \N(g\circ U)} \non \\
      &=& {1\over V} \left[ \sum_\vx  | \overline{\phi}(\vx;g\circ U,g\circ \eta)|\right]_{\eta \in \N(U)} \non \\
      &=& {1\over V} \left[  \sum_\vx  | g(\vx) \overline{\phi}(\vx;U,\eta)|\right]_{\eta \in \N(U)}  \non \\
      &=&  \Phi(U) \ ,
\label{Pginv}
\eea
which establishes the gauge invariance of the spin glass order parameter. 
The term proportional to $h$ serves exactly the same function as in any spin model with a global symmetry; i.e.\ it breaks the global symmetry explicitly.  Without this term,  $\overline{\phi}(\vx;U,\eta)$ (which is evaluated at fixed $U$) would 
vanish in a finite volume, due to the custodial symmetry of  $H(\phi,U)$, as would a spin $s_x$ in the Ising model in the absence of an external field, due to the global $Z_2$ symmetry.  But this breaking term does {\it not} also
break gauge invariance.  The order parameter, as we have just seen, is gauge invariant, even at finite $h$.  As in any spin model, the $h \ra 0$ limit follows the infinite volume limit.   

   In the Edwards-Anderson model \rf{EA} at $h\ra 0$ there are a vast number of configurations which are nearly degenerate in energy, and there will be an analogous phenomenon in the spin glass phase of the gauge Higgs theory.  We defer a discussion of this point to section \ref{Gribov}.
   
   From \rf{ZsR} and point 1 above, we see that
\bea
          \Zs(U) &\equiv& \Zs(U,\eta)_ {|\eta \in \N(U)}
\eea
is independent of which element $\eta$ is chosen in the set $\N(U)$, and is also gauge invariant:
\bea
          \Zs(g\circ U) &=& \Zs(g\circ U,\eta')_{|\eta' \in \N(g\circ U)}\non \\
                               &=& \Zs(g \circ U, g\circ \eta)_{|\eta \in \N(U)} \non \\
                               &=& \Zs(U,\eta)_{|\eta \in \N(U)} \non \\
                               &=& \Zs(U) \ ,
\eea
where we have used \rf{Zsg} and point 2.  We then define
\bea
          Z &=& \int DU_i(\vx) \ \Zs(U) \non \\
             &=& \int DU_i(\vx) D\phi(\vx) \ e^{-\Hs(\phi,U,\eta\in \N(U))/kT}  \ .
\eea

   In a spin glass, $P(J)$ can be taken as the product of probability distributions for each $J_{ij}$, which are typically
taken to be Gaussian distributions $\exp(-J_{ij}^2/2J^2)$, or else $J_{ij} = \pm J$ with equal probability for each sign,
and the pairs of sites $i,j$ are sometimes chosen to be nearest neighbors.  In gauge Higgs theory, however, $P(U)$ is determined from the condition that the expectation value of a gauge invariant operator $Q(U)$ that depends only on $U$ is given by 
\bea
            \langle Q \rangle &=& {\tr \ Q e^{-\Hs/kT} \over \tr \ e^{-\Hs/kT}} \non \\            
                                       &=& {1\over Z} \int DU_i(\vx) Q(U) \int D\phi(\vx) 
                                                    e^{-\Hs(\phi,U,\eta \in \N(U))/kT}  \non \\
                                        &=&  {1\over Z} \int DU_i(\vx) \ Q(U) \Zs(U) \non \\
                                        &=&  \int DU_i(\vx) \ Q(U) P(U) \ .
\eea
Therefore $P(U)$ is the gauge invariant probability density  \footnote{In 1978 Hertz \cite{Hertz:1978zza} put forward a spin glass version of an abelian gauge theory with
matter.  It differs from ours precisely in the choice of $P(U)$, which in \cite{Hertz:1978zza} was taken to be the usual Boltzmann factor of a {\it pure} gauge theory in $D=4$ Euclidean dimensions.  This means that expectation values of gauge invariant quantities, in such a spin glass version of gauge theory, differ from that of gauge Higgs theory; these
are different theories.  In our formulation, with $P(U)$ given by \rf{PU}, the point is that the standard theory is already a spin glass theory.}
\beq
             P(U) = {\Zs(U) \over Z} \ .
\label{PU}
\eeq
With this probability density, $\langle Q \rangle$ is the standard expectation value of $Q(U)$ in a gauge Higgs theory in the $h=0$ limit.  This completes the definition of $\Phi(U)$ and $\langle \Phi \rangle$, and the proof of gauge invariance.

We now have a gauge invariant criterion for the spontaneous breaking of custodial symmetry:
\bea
           \lim_{h \ra 0} \lim_{V \ra 0} \langle \Phi \rangle \left\{ \begin{array}{cl}
                           = 0 & \text{unbroken symmetry} \cr
                          > 0  & \text{broken symmetry} \end{array} \right. \ ,
\label{criterion}
\eea
which is entirely analogous to the Edwards-Anderson spin glass criterion
\bea
           \lim_{h \ra 0} \lim_{V \ra 0} \langle q \rangle \left\{ \begin{array}{cl}
                           = 0 & \text{non-spin glass phase} \cr
                          > 0  & \text{spin glass phase} \end{array} \right. \ .
\eea
There has always been a question, in SU(2) and other gauge Higgs theories with the Higgs field in the fundamental
representation, of how to distinguish the Higgs phase from the region with QCD-like physics in the absence of
a thermodynamic transition.  Our suggestion
is that the Higgs phase is the phase of broken custodial symmetry, as defined by the criterion stated above.  As such,
the Higgs phase is the spin glass phase of the gauge Higgs theory.  Our task is to show that the distinction in terms of
symmetry corresponds to a physical distinction between the spin glass phase and the QCD-like phase of a gauge Higgs theory, in terms of the type of confinement present in each phase.  This will be deferred to section \ref{coincide}.

\section{\label{numbers} Numerical evaluation}

    The Edwards-Anderson Hamiltonian $\Hs(\{s_x\},\{J_{ij}\})$ for the Ising spin glass system is a simple expression, easily calculated for any given spin configuration.  The same cannot be said for $H(\phi,U)$ of the gauge Higgs theory, defined in eq.\  \rf{H},
for which we do not have an 
explicit form.  Fortunately, by the usual arguments, $H(\phi,U)$ can be expressed in terms of a Euclidean time path integral, which makes the computation of $\langle \Phi \rangle$ amenable to lattice Monte Carlo methods.  Identifying the arguments of $H$, i.e.\ $\phi(\vx),U_i(\vx)$, as the
Euclidean time dependent fields $\phi(\vx,0), U_i(\vx.0)$ on the $t=0$ time slice, we have
\bea
      & &     \exp[-H(\phi(\vx),U_i(\vx))/kT]  \non \\ 
     & & \ = \int DU_0 [DU_i D\phi]_{t\ne 0}  \exp[-S_E(\phi(\vx,t),U_\m(\vx,t))] \ , \non \\
\label{euclid}
\eea
where $S_E$ is the Euclidean action.  The notation $[DU_i D\phi]]_{t\ne0}$ means that only fields at times $t\ne 0$ are integrated over.\footnote{The restriction does not apply to $U_0$ which can, if desired, be fixed to ${U_0=\mathbbm{1}}$ everywhere except on a single time slice on the periodic lattice.} Periodic boundary conditions, and a lattice time interval 
${-\oh N_t \le t <\oh N_t}$, where $N_t = 1/(kT a)$ with $a$ the lattice spacing, are understood.  The invariance of $H$
under custodial  transformations $H(\phi(\vx)R,U_i(\vx))=H(\phi(\vx),U_i(\vx))$ and gauge transformations is derived from
the invariance of $S_E$ under these transformations, e.g.\ defining, at $t\ne 0$, ${\phi'(x,t) = \phi(x,t) R}$,
\begin{widetext}
\bea
         \exp[-H(\phi(\vx)R,U_i(\vx))/kT] 
  &=& \int DU_0 [DU_i D\phi']_{t\ne 0}   \exp[-S_E(\phi(\vx,0)R,\phi'(\vx,t\ne0),U_\m(\vx,t))] \non \\
  &=& \int DU_0 [DU_i D\phi]_{t\ne 0}  \exp[-S_E(\phi(\vx,0)R,\phi(\vx,t\ne0)R,U_\m(\vx,t))]  \non \\
     &=& \int DU_0 [DU_i D\phi]_{t\ne 0}  \exp[-S_E(\phi(\vx,0),\phi(\vx,t\ne0),U_\m(\vx,t))]  \non \\
     &=&    \exp[-H(\phi(\vx),U_i(\vx))/kT] \ ,
\label{Hcust}
\eea
\end{widetext}
which demonstrates the invariance of $H$ under custodial transformations.  Similar manipulations show that $H$ is gauge invariant.
\begin{figure*}[t!]
\subfigure[~]  
{   
 \label{extrap}
 \includegraphics[scale=0.65]{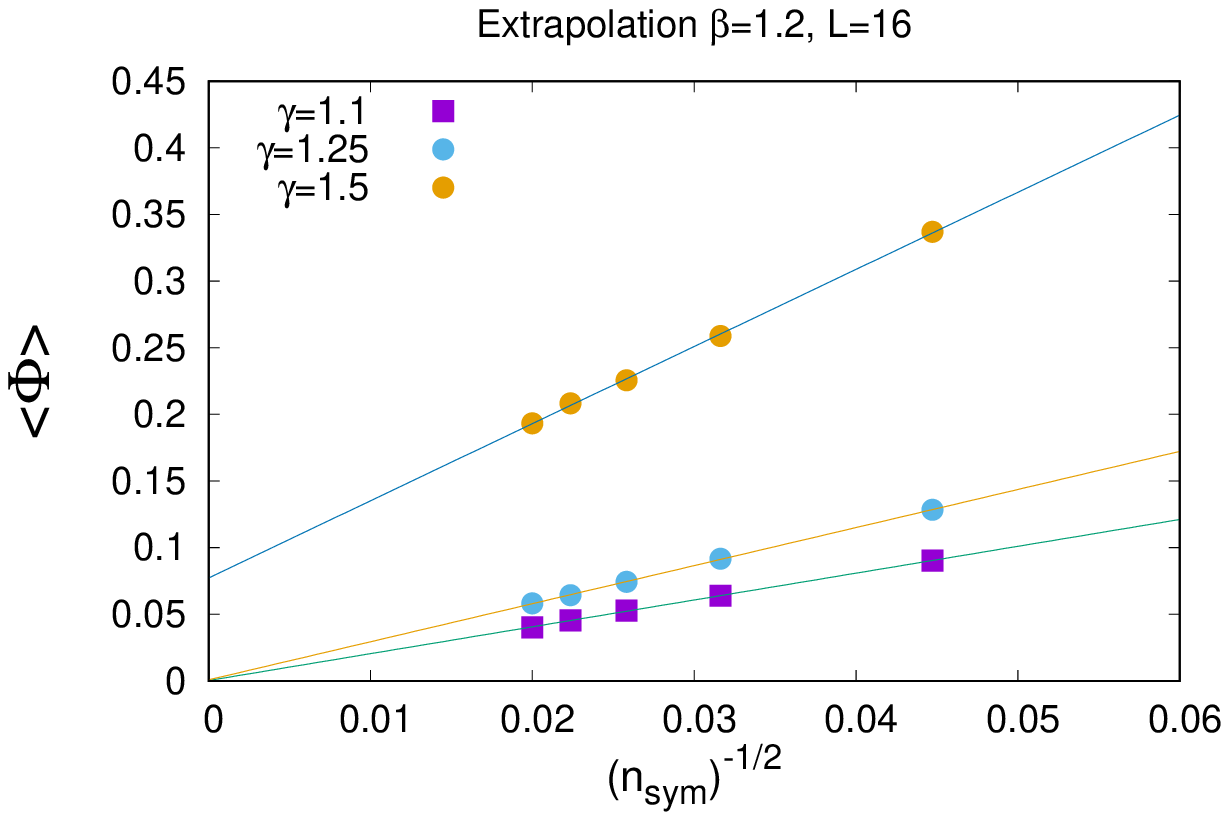}
}
\subfigure[~]  
{   
 \label{phase}
 \includegraphics[scale=0.65]{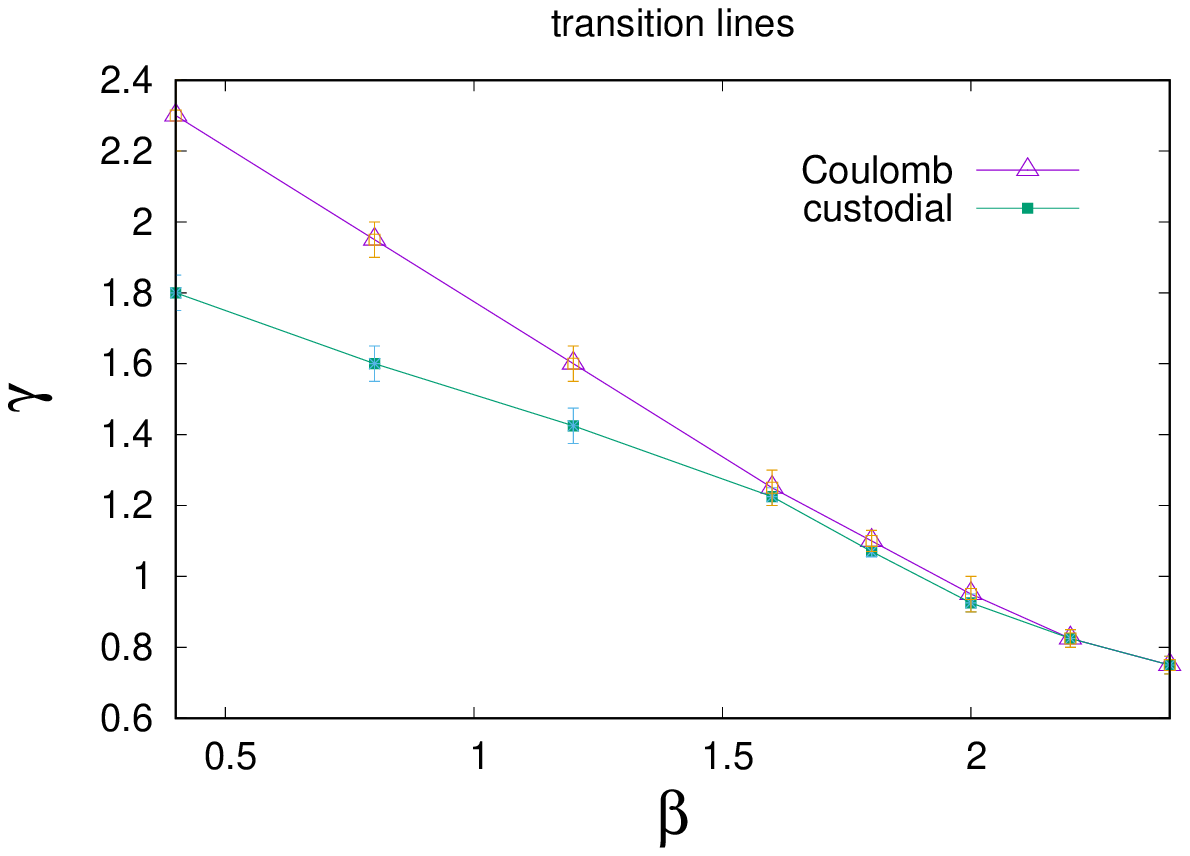}
}
\caption{(a) Extrapolation of $\langle \Phi \rangle$ to $n_{sym}\ra \infty$ above ($\g=1.5$) and below ($\gamma=1.1,1.25$) the custodial symmetry breaking transition at $\b=1.2,\g=1.4$, in SU(2) gauge Higgs theory. The lattice volume is $16^4$; error bars are smaller than the symbol sizes.  (b) Transition lines separating the unbroken (smaller $\g$) and broken (larger $\g$) phases of (i) remnant gauge symmetry in Coulomb gauge (triangular data points), and (ii) custodial symmetry (filled circles).  Note that the broken Coulomb phase lies entirely
within the broken custodial symmetry phase, as it must from the bound in \rf{bound}.} 
\label{data}
\end{figure*}

   In the numerical computation of $\ophi(\vx;U,\eta)$ it is permissible to go immediately to the $h=0$ limit.\footnote{Where $h$ should not be confused with Planck's constant.}  We will need
$h \ne 0$ to prove formal identities in the next section, but it is of no importance in the Monte Carlo simulation.
For those
simulations we drop $\eta$, and denote our observable as just $\ophi(\vx,U)$.  
The computational procedure is to (i) generate a set of uncorrelated $U_i(\vx)$ configurations drawn from the probability distribution $P(U)$; and (ii) evaluate $\Phi(U)$ in each configuration, finally averaging the resulting set of $\Phi(U)$ to estimate $\langle \Phi \rangle$.  This amounts to a 
Monte-Carlo-within-a-Monte-Carlo simulation.  The set of $U_i(\vx)$ configurations are obtained by running the usual Monte Carlo simulation of the Euclidean lattice, updating both the link variables and scalar field together on the full lattice volume.
After a sufficient number of update sweeps of this kind (``sufficient'' means that the final configuration is uncorrelated with
the initial configuration), the $U_i(\vx,0)$ configuration on a $t=0$ time slice (or, for that matter, on any time slice) is
obviously drawn from the probability distribution $P(U)$, since a set of gauge field configurations generated in this way would
give the correct expectation value $\langle Q \rangle$ of any gauge invariant observable that depends only on the spatial
link variables on a time slice.  With a configuration $U_i(\vx,0)$ in hand, and taking the existing 
$U(\vx,t),\phi(\vx,t)$ configuration on the Euclidean time lattice as the initial configuration, we then compute 
$\ophi(\vx,U)$ from
a Monte Carlo simulation of the Euclidean action $S_E$, with $U_i(\vx,0)$ on the $t=0$ time slice fixed, i.e.
\beq
            \ophi(\vx,U) = {1\over Z_{spin}(U)} \int DU_0 [DU_i]_{t\ne 0} D\phi ~ \phi(\vx,t=0) ~ e^{-S_E} \ .
\eeq    

    So custodial symmetry breaking, or equivalently the spin glass transition in gauge Higgs theory, is determined as follows:  The SU(2) gauge and scalar fields are updated in the usual way, but each data-taking sweep actually consists of a set of $n_{sym}$ sweeps in which the spacelike links $U_i(\vx,0)$ are held fixed on the $t=0$ time slice.  Let $\phi(\vx,t=0,n)$ be the scalar field at site $\vx$ on the $t=0$ time slice at the $n$-th sweep.  Then we compute $\ophi(\vx,U)$ from the average over $n_{sym}$ sweeps 
\beq
             \ophi(\vx,U) = {1\over n_{sym}} \sum_{n=1}^{n_{sym}} \phi(\vx,0,n) \ ,
\eeq
and  the order parameter $\Phi(n_{sym},U)$ from \rf{max}.  Here it is important to indicate the dependence
on $n_{sym}$. Then the procedure is repeated, updating links and the scalar field together, followed by another computation of $\Phi(n_{sym},U)$ from a simulation with spatial links at $t=0$ held fixed, and so on.     Averaging the $\Phi(n_{sym},U)$ obtained by these means results in an estimate for $\langle \Phi(n_{sym}) \rangle$.  Since $\Phi(n_{sym},U)$ is a sum
of moduli, it cannot be zero.  Instead, on general statistical grounds, we expect~\footnote{One must keep in mind that 
at finite $V$, $ \langle \Phi \rangle$ would actually vanish at ${n_{sym} \ra \infty}$, since a symmetry cannot break in a finite volume.  The proper order of limits is first $V \ra \infty$, then $n_{sym} \ra \infty$.  Nevertheless, for $n_{sym}$ not too large,  \rf{fit} is a good fit to the data, and the extrapolation should be reliable.} 
\beq
           \langle \Phi(n_{sym}) \rangle = \langle \Phi \rangle + {\k \over \sqrt{n_{sym}}} \ ,
\label{fit}
\eeq
where $\k$ is some constant.  
By computing  $\langle \Phi(n_{sym}) \rangle$ in independent runs at a range of $n_{sym}$ values, and fitting
the results to \rf{fit}, we obtain an estimate for $\langle \Phi \rangle$ at any point in the  $\b,\g$ plane of lattice couplings,
and temperature $T$.  

   We will be mainly interested in the phase diagram in the plane of lattice couplings at zero temperature, which means using a lattice with a sufficiently large extension in the Euclidean time direction to approximate $T=0$.  This is, of course, a departure from the Edwards-Anderson spin glass, where one is instead interested in the transition at finite temperature.  The transition points at (approximately) zero temperature in the SU(2) gauge Higgs model are determined by varying $\g$ at fixed lattice coupling $\b$.  At values of $\g$ below the spin glass/custodial symmetry breaking transition, the data for  
$\langle \Phi(n_{sym}) \rangle$ extrapolates to $\langle \Phi \rangle = 0$ as $n_{sym}\ra \infty$.  Above the transition, this data extrapolates to a finite value.  Transition points are estimates of where the extrapolated $\langle \Phi \rangle$ value begins to move away from zero, as $\g$ increases.  An example of the data below and above the transition, at fixed
$\b=1.2$, is displayed in Fig.\ \ref{extrap}.  The custodial symmetry breaking transition line, joining transition points determined as just described, is shown in Fig.\ \ref{phase}.  

   It is useful to compare the spin glass/custodial symmetry breaking line with the transition line for remnant gauge symmetry breaking in Coulomb gauge, which is also displayed in Fig.\ \ref{phase}.  The Coulomb gauge fixing condition
leaves unfixed a remnant symmetry $g(\vx,t) = g(t) \in$ SU(2) which is global on any time slice.  Then at each data taking sweep 
we fix to Coulomb gauge, and define
\beq   
      |\phi_{av}(t)| = \left|{1\over V} \sum_{\vx} \phi(\vx,t) \right| \ ,
\eeq
where $V$ is still the three-volume of a time slice, with susceptability
\beq
     \chi =  {1\over N_t} \sum_{t=1}^{N_t} V ( \langle |\phi_{av}(t)|^2 \rangle_C - \langle |\phi_{av}(t)| \rangle_C^2 ) \ ,
\eeq
where the subscript $C$ means that the observable is evaluated in Coulomb gauge. The remnant symmetry transition points shown in Fig.\ \ref{phase} are estimated from peaks in the susceptibility.
Note that these points lie above the custodial symmetry breaking/spin glass transition line.  The reason for this 
will be explained in section \ref{gauge}.

\subsection{\label{Gribov} Gribov copies and the spin glass phase}

    The gauge fixing sweeps that are used to fix to a gauge in lattice Monte Carlo, and the sweeps at fixed $U_i(\vx,0)$ used
 to compute $\Phi(U)$ in the gauge Higgs theory, have something in common.  Gauge fixing sweeps in, e.g., Coulomb gauge seek to maximize the quantity
\beq
           R = \sum_x \sum_{i=1}^3 \text{Re}[\tr U_i(x)] \ .
\eeq
But in practice no gauge fixing algorithm exists which can fix to an absolute maximum of $R$.   The problem is analogous to finding the spin configuration for which the spin glass Hamiltonian is an absolute minimum in the $h\ra 0$ limit.  The Hamiltonian $H_{spin}$  in \rf{EA} has a very large number of near-degenerate minima, and the global minimum is impossible to determine in practice. In the gauge fixing case, the best that can be done is to fix to one of a vast number of local maxima, which are the Gribov copies.  Computer algorithms are deterministic, and reach a unique local maximum on the gauge orbit, but which maximum is obtained depends on the starting configuration on the gauge orbit. 

    In the Edwards-Anderson model in the spin glass phase, the spins fluctuate around one of the near-degenerate minima, which is in general not the global minimum of the energy.  This same phenomenon is also seen in gauge Higgs theory, when calculating the order parameter $\Phi(U)$ from the Monte-Carlo-within-a-Monte-Carlo procedure.   In the spin glass phase of gauge Higgs theory, in the data-taking part of the simulation, the scalar field on the $t=0$ time slice fluctuates around some configuration, dependent on the starting configuration, with non-vanishing $\Phi(U)$.  As in the Edwards-Anderson model there are a vast number of such metastable configurations in the spin glass phase, for fixed $U_i(\vx,0)$, which give rise to non-zero but slightly different $\Phi$.  In practice we find that the statistical error in $\Phi$ is on the order of one or two percent, so clearly these stable (or, in a finite volume, metastable) configurations have very nearly the same value for the order parameter.   In spin glass theory the degenerate configurations are believed to be thermodynamically equivalent.
Which particular configuration, in the data taking sweeps of a gauge Higgs theory in the spin glass phase, happens to be singled out by the starting configuration is likely to be of little physical importance. \\

\section{\label{gauge} Custodial and gauge symmetry breaking}

   We know from the Elitzur theorem that a local gauge symmetry cannot break spontaneously.  Nevertheless,
if we impose a physical (e.g.\ Coulomb or axial) gauge which leaves a global remnant symmetry on a time slice,
then it is possible that the remnant gauge symmetry can break on that time slice.  We will now show that
custodial symmetry breaking is a necessary condition for remnant gauge symmetry breaking in any physical gauge, and a sufficient condition for the existence of remnant symmetry breaking in some physical gauge.  This is a prerequisite to
our following discussion of  \Sc \ and C confinement.
 
    A physical gauge refers to a gauge in which there exists a ghost free Hamiltonian; this excludes gauge conditions which couple link variables on different time slices (as in lattice Landau gauge).  We will consider physical gauges specified by conditions of the form $F(U)=0$, where the condition is imposed on spacelike link variables on each time slice, as in lattice Coulomb and axial gauge, removing all local gauge symmetry (but leaving some remnant global symmetry) on a given time slice.   We will refer to gauges of this type as ``$F$-gauges.''  Since $\Phi(U)$ is gauge invariant, it can of course be evaluated in any 
particular gauge, i.e.
\begin{widetext}
\bea
 \langle \Phi \rangle &=&  \int DU \delta[F(U)] \D_F[U] \Phi(U) P(U) \non \\
  &=&  {1\over Z} \int DU \delta[F(U)] \D_F[U]   {1\over V} 
      \max_\eta \sum_\vx \left| \int D\phi \phi(x)  e^{-(H(\phi,U) 
  -  h\sum_\vx \tr \eta^\dg(x)\phi(\vx))/kT} \right|  \ , \non \\
 \eea
 where $\D_F(U)$ is the Faddeev-Popov determinant. It should be noted that in lattice gauge-fixing algorithms,
 every given lattice configuration is transformed deterministically to a single 
 gauge-fixed configuration with $\D(U)>0$.  This is how lattice simulations evade Neuberger's theorem 
 \cite{Neuberger:1986xz}. In lattice
Monte Carlo simulations  the gauge-fixing algorithm makes a choice among gauge copies, and should probably be regarded as part of the specification of the gauge choice.

The modulus of the scalar field expectation value in an $F$-gauge is 
\bea
      |\langle \phi \rangle_F| &=& \lim_{h \ra 0} \lim_{V\ra \infty} \left| {1\over Z} \int DU \delta[F(U)] \D_F[U] 
       {1\over V} \sum_\vx  \int D\phi \, \phi(x) e^{-(H(\phi,U) 
  -  h\sum_\vx \tr \phi(\vx))/kT} \right|\non \\
      &\le& \lim_{h\ra 0} \lim_{V\ra \infty} {1\over Z} \int DU  \delta[F(U)] \D_F[U]  
       {1\over V} \sum_\vx \left| \int D\phi \, \phi(x) e^{-(H(\phi,U) 
  -  h\sum_\vx \tr \phi(\vx))/kT} \right| \non \\
      &\le& \lim_{h \ra 0} \lim_{V\ra \infty} {1\over Z} \int DU \delta[F(U)] \D_F[U]   
        {1\over V} \max_\eta \sum_\vx \left| \int D\phi \, \phi(x)  e^{-(H(\phi,U) 
               - h\sum_\vx \tr \eta^\dg(x)\phi(\vx))/kT} \right| \non \\
      &\le&  \lim_{h \ra 0} \lim_{V\ra \infty} \int DU \delta[F(U)] \D_F[U]  {\Zs(U) \over Z} \Phi(U) \non \\
      &\le& \langle \Phi \rangle \ .
\label{bound}
\eea
\end{widetext}
Equation \rf{bound} means that spontaneous breaking of custodial symmetry, $\langle \Phi \rangle > 0$ in the thermodynamic limit, is a necessary condition for the spontaneous breaking of a remnant 
gauge symmetry in any physical $F$-gauge.\footnote{Note that we have not distinguished in \rf{bound} between partition functions $Z$ with different symmetry breaking terms proportional to $h$, since their ratios equal unity in the limits shown.}

     Since custodial symmetry is a continuous symmetry, one might expect Goldstone modes in the broken phase of custodial
symmetry, resulting in long range correlations among the $\phi$ fields in certain Green's functions at fixed $U$.  Such long range correlations are, however, gauge dependent, and vanish when integrating over $U$.  Moreover, one of the assumptions of the Goldstone theorem is that there are only short range couplings in the Hamiltonian.  In general this assumption is violated in a physical gauge that removes all local gauge symmetry, as pointed out long ago by Guralnik et al.\ \cite{Guralnik:1967zz}.  For these reasons, spontaneous breaking of custodial symmetry and/or remnant gauge symmetry in a physical gauge are not associated with massless Goldstone particles.

    Next let $\teta(\vx;U) \in \N(U)$ be a choice of one member from each set $\N(U)$, with 
$\teta(\vx;g\circ U) = g(\vx) \teta(\vx;U)$; this choice is possible for reasons noted below \rf{equiv}.  
Then, from \rf{ophig}, $\ophi(\vx;U,\teta(U))$ is a gauge covariant functional of $U$. We define the gauge $\hF(U)=0$ as the condition
\beq
           \hF(U) \equiv {\ophi(\vx;U,\teta) \over |\ophi(\vx;U,\teta)|} - \mathbbm{1} = 0 
\label{Fhat}
\eeq
at all $\vx$ on the time slice.  In this gauge
\begin{widetext}
\bea
\langle \Phi \rangle &=& \lim_{h \ra 0} \lim_{V\ra \infty}  \int DU \d[\hF(U)] \D_{\hF}[U] 
\left\{ {1\over V} \sum_\vx |\ophi(\vx;U,\teta)| \right\}{Z_{spin}(U)\over Z} \label{equal2} \\
&=& \lim_{h \ra 0} \lim_{V\ra \infty}  \int DU \d[\hF(U)] \D_{\hF}[U] 
   \left\{ {1\over V}\left| \sum_\vx \ophi(\vx;U,\teta) \right| \right\} {Z_{spin}(U)\over Z} \label{equal3} \\
    &=& \lim_{h \ra 0} \lim_{V\ra \infty} \left|  {1\over V}   \sum_\vx {1\over Z}  \int DU \d[\hF(U)] \D_{\hF}[U]   
           \int D\phi \ \phi(\vx) e^{-(H(\phi,U)-h\sum_\vx \tr \teta^\dg(\vx;U) \phi(\vx))/kT}  \right| \label{equal4} \\
     &=& |\langle \phi \rangle_{\hF}| \ ,
\label{equal}
\eea
\end{widetext}
where in passing from \rf{equal2} to \rf{equal3} we make use of \rf{Fhat}.
 
Broken custodial symmetry is therefore also a sufficient condition for the existence of a physical $F$-gauge
in which the expectation value $\langle \phi \rangle_F$ is non-zero.  We note that in the $h\ra 0, V \ra \infty$ limits the
details of the symmetry breaking term in the computation of $\langle \phi \rangle_{\hF}$ should not be important, and for
 $\langle \phi \rangle_{\hF} \propto \mathbbm{1}$ any term which biases $\phi$ slightly towards the identity matrix should suffice.  In particular $\teta$ could be replaced by $\mathbbm{1}$ in \rf{equal4} without affecting $\langle \phi \rangle_{\hF}$ in the appropriate limits.

\section{\label{variety} C and \Sc \ Confinement}
    
       In a gauge Higgs theory with the matter field in the fundamental representation of the gauge group, as in QCD, large Wilson loops have a perimeter-law falloff, Polyakov loops have a finite expectation value, so in what sense are these theories confining?  The usual
answer is that confinement simply means that the asymptotic particle spectrum is color neutral, which in turn means that
such particles are not the sources of a gauge field that could be detected far from the source.  This property is often
called ``color confinement;''   we will refer to it as ``C confinement'' for short.  Note that in a gauge Higgs theory, where there is no thermodynamic separation between the QCD-like and Higgs regions of the phase diagram, the property of C confinement holds (in $D\le 4$ dimensions) throughout the phase diagram, including deep in the Higgs regime.  In the 
abelian Higgs model in $D=4$ dimensions with a compact U(1) gauge group and a single charged scalar field, C confinement holds everywhere outside the massless Coulomb phase.

     In a pure gauge theory, however, there exists a variety of confinement which is stronger than color confinement, which we will call ``separation of charge confinement'' or  ``\Sc \ confinement.''  Certainly C confinement holds true in a pure gauge theory,  whose particle spectrum consists of color neutral glueballs.  What distinguishes the pure gauge theory from
a gauge theory with matter in the fundamental representation is the existence of a confining static quark potential.
Let $q,\q$ be static quark/antiquark operators, and define
\beq
          Q(R) = \q^a(\vx)  V^{ab}(\vx,\vy;U) q^b(\vy) \ ,
\label{VU}
\eeq    
where $V$ is an operator which is a functional of the spacelike link variables $U_i$ and which transforms, under a gauge transformation, like a Wilson line running between points $x_1$ and $x_2$, and $R=|\vy-\vx|$.  Contraction of the Dirac indices is implicit. We consider gauge invariant states containing these static quark-antiquark sources by letting $Q(R)$ operate on the vacuum, i.e.
\beq
        \Psi_V(R) = \q^a(\vx)  V^{ab}(\vx,\vy;U) q^b(\vy) \Psi_0 \ ,
\label{PsiV}
\eeq
and it is convenient to normalize $V$ to agree with the normalization of a Wilson line, i.e.
\beq
           \langle \Psi_0| \tr [V^\dg(\vx,\vy;U) V(\vx,\vy;U)]|\Psi_0 \rangle = N \ ,
\label{norm}
\eeq
where $N$ is the number of colors.
It is not hard to see that the energy expectation value of this state above the vacuum energy is obtained from the logarithmic time derivative
\bea
      E_V(R) &=& - \lim_{\e\ra 0}{d\over d\e} \log \left[ \langle [Q^\dg(R)]_{t=+\oh \e} [Q(R]_{t=-\oh \e} \rangle \right] \non \\
                   &=&   {1 \over N} \langle \Psi_V|(H-E_0)| \Psi_V \rangle \ ,  \non \\
\label{logtime}
\eea
where $E_0$ is the vacuum energy, and in the first line the expectation value is evaluated in a Euclidean path integral with a large extension in the Euclidean time direction.  The notation $[..]_t$ means that the operator is applied at time $t$.    The minimum possible energy $E_{min}(R)$ is the static quark potential, as determined from the behavior of large Wilson loops.  Since, in a pure gauge theory, $E_{min}(R) \sim \s R$ at large $R$, and $E_V(R)$ is bounded from below by $E_{min}(R)$, 
it follows that
\beq
          \lim_{R \ra \infty} E_V(R) = \infty  ~~~ \text{for all~~} V(\vx,\vy;U) \text{~~operators} \ .
\label{Sc}
\eeq
We will refer to this property as ``separation of charge'' (\Sc) confinement.  It is a stronger condition than C confinement,
and the question is whether this definition can be extended to gauge theories with matter in the fundamental representation.

    Our proposal in \cite{Greensite:2017ajx} is simple:  eq.\ \rf{Sc} is {\it also} the definition of \Sc \ confinement in gauge Higgs theories, and other gauge + matter theories.  The crucial condition is that the operator $V(\vx,\vy;U)$ depends only on the spacelike link variables, and not on the matter fields.  If we imagine taking the $\vy \ra \infty$ limit in \rf{VU}, then the physical
state $\Psi_V$ represents an isolated quark at point $\vx$, together with a surrounding color electric field so as to satisfy the Gauss law, and the question is whether such states, in a gauge Higgs theory, can ever be of finite energy for some choice
of $V$ depending only on $U$.  In contrast, if $V$ is allowed to involve matter fields, we could construct operators such as, e.g.
\bea
   V^{ab}(\vx,\vy;\phi) &=& \phi^a(\vx) \phi^{\dg b}(\vy) \non \\
   Q(R) &=&  \q^a(\vx) \phi^a(\vx) \phi^{\dg b}(\vy) q^b(\vy) \ ,
\label{Vp}
\eea
which create two color singlet quark-scalar systems, localized at points $\vx, \vy$, with a
negligible $R$-dependent interaction energy.  States of that type would be obtained after string breaking, and we therefore exclude such operators
from the \Sc \ criterion.  That is not to say that states created by the operators \rf{Vp} are completely orthogonal
to $\Psi_V(R)$ states, but they may become orthogonal in the $R\ra \infty$ limit; that is also a question of interest which
we address in the next section.  

Note that it is always possible to find $V(\vx,\vy;U)$ operators for which $E_V(R)$ diverges with $R$.  A simple Wilson line running between the quark-antiquark sources is an example, and in fact such a state has an energy which rises linearly with $R$ even in an abelian, non-confining theory.  The \Sc \ criterion is that $E_V(R)$ diverges at $R\ra \infty$ for all $V$, regardless of whether $\Psi_V(R)$ evolves, in Euclidean time, towards a ``broken string'' state.  But if there is any $V$ operator which violates the \Sc \ criterion, then, assuming the absence of a massless phase, the system is in a C, rather than an \Sc, confining phase. 
   
   Let us also note in passing that \Sc \ confinement requires that the gauge group has a non-trivial center.  If the center is
trivial then it is always possible to construct local operators $\xi(\vx;U)$ which depend only on the gauge field, and
which transform in the fundamental representation of the gauge group.  In that case one could construct operators as
in \rf{Vp} by replacing $\phi$ with $\xi$.  These operators again create two color neutral objects, invariant under all transformations in the gauge group, whose interaction energy is negligible at large separation.   

   To investigate these matters in a regulated, non-perturbative formulation, amenable to numerical simulation, we must go to
the Euclidian lattice formulation, and replace the continuous time derivative by its discretized version.  
After integrating out the static quark antiquark fields and dropping an $R$-independent mass term, the result is  
\begin{widetext}
\bea
 E_V(R) &=& - \log \left[ {1\over N} \langle \tr [ U^\dg_0(\vx,t)  V(\vx,\vy,t;U) U_0(\vy,t) V^\dg(\vx,\vy,t+1,U)] \rangle  \right] \  ,
\label{Elat}
\eea
\end{widetext}
where it is understood that $V(\vx,\vy,t,U)$ depends only on the spacelike link variables $U_i(\vx,t)$ on time slice $t$,
and the \Sc \ criterion applies to this lattice version of $E_V(R)$.\footnote{In making the lattice approximation \rf{Elat} to the logarithmic time derivative \rf{logtime} in the \Sc \ phase, the lattice spacing in the time direction must be taken small compared to the Eucidean time required for $\Psi_V(R)$ to evolve to a state containing two isolated quark-scalar singlets, with an $R$-independent energy expectation value.}

The first question is whether a gauge Higgs theory is \Sc \ confining {\it anywhere} in the $\b,\g$ coupling plane, apart from the pure gauge theory at $\g=0$, and the answer is yes.  We have shown that the \Sc \ condition is satisfied at least in the strong coupling region, using the lattice strong coupling expansion \cite{Greensite:2018mhh}. Then the second question is whether the \Sc \ criterion is 
obeyed {\it everywhere} in the phase diagram, and the answer is no.  In \cite{Greensite:2017ajx} we showed that in some region of the lattice SU(2) gauge Higgs phase diagram there are $V$ operators that can be inserted in \rf{Elat} which violate the \Sc \ condition, and in that region the theory is C but not \Sc \ confining.  This means that there must exist a transition of some kind between the \Sc \ and C confinement regions.  The question is where that transition occurs, and whether it coincides with the custodial symmetry/spin glass transition.

\section{\label{coincide} Coincidence of the spin glass and confinement transitions}

   We now show that the spin glass phase is a Higgs phase, i.e.\ it is a phase in which a global subgroup
of the gauge group is broken spontaneously, and as such it is a C confining phase.   To justify this statement
we will consider quantizing a gauge Higgs theory in the $F$-gauges, in which the field operators $\phi,q,\q$, acting
on the vacuum, create physical states.  The unitary, covariant, and temporal gauges will
be discussed shortly.
 
   Any physical $F$-gauge leaves unfixed a global subgroup $G$ of the gauge group  which preserves the gauge condition.  At a minimum this includes the center of the gauge group, so the remnant gauge symmetry includes at least the 
transformations
\beq
      \phi(\vx,t) \ra z(t) \phi(\vx,t) ~~~,~~~ U_0(\vx,t) \ra z(t) U_0(\vx,t) z^\dg(t+1) \ . \\
\label{remnant}
\eeq
Some gauges, e.g.\ Coulomb gauge, preserve a larger remnant symmetry under gauge transformations 
$g(\vx,t)=g(t) \in$ SU(N).  Other gauges, e.g.\ some versions of axial gauge, which preserve only those transformations $g(t)$ which are diagonal matrices, are more restrictive.   But any $F$-gauge preserves at least the symmetry under the transformations \rf{remnant}, and there are examples which preserve only that global symmetry, as discussed below.

      The field operators $\phi, q, \overline{q}$ transform under the remnant gauge symmetry in an $F$-gauge and, acting
on the vacuum, create physical states.   These operators do not only excite the vacuum state at one point $\vx$, but must also,
in accordance with the Gauss law, create a color electric field..  While this associated color electric field may carry only a finite amount of energy, it may also be of infinite energy.

Let $g_F(\vx;U)$ be the gauge transformation which takes the gauge field $U$ into the $F$-gauge.\footnote{It is assumed that
$g_F(\vx;U)$ is uniquely determined from $U$ by the gauge-fixing algorithm, cf.\ section \ref{Gribov}, although different 
configurations on a gauge orbit may be transformed into different Gribov copies}  Denoting field operators in an $F$-gauge by the subscript $F$, they have the form
\bea
          \phi_F(\vx) &=& g_F(\vx,U) \phi(\vx) \non \\
          q_F(\vx) &=& g_F(\vx,U) q(\vx) \non \\
          \overline{q}_F(\vx) &=& \overline{q}(\vx) g^\dg_F(\vx,U) \non \\
          U_{i,F}(\vx) &=& g_F(\vx,U) U_{i}(\vx) g^\dg_F(\vx+\hat{\i},U) \ .
\label{F}
\eea
The $\phi_F, q_F, \overline{q}_F$ field operators are invariant under local gauge transformations, but
still transform under global transformations in the remnant gauge group $G$.  At a minimum, these operators transform 
under global gauge transformations ${g(\vx) = z \in Z_N}$.
A well known example, in continuum abelian gauge theory, is the Coulomb gauge operator
\beq
         \phi_C(\vx) = g_C(\vx;A) \phi(\vx) \ ,
\label{dress}
\eeq
where
\bea
            g_C(\vx;A) &=& \exp\left[i {e\over 4\pi} \int d^3z ~ A_i(\vz) {\pa \over \pa z_i} {1\over |\vx -\vz|} \right]  \ .
\label{Dirac}
\eea
One can check that $g_C(\vx;A)$ is the gauge transformation to Coulomb gauge, and also that it creates the Coulomb
electric field associated with a static charge at point $\vx$.  In an abelian theory in any $F$-gauge, such as Coulomb
gauge, the remnant gauge symmetry is $g(\vx) = e^{i\th}$.  Under an arbitrary gauge transformation $g(\vx) = e^{i\th(\vx)}$,
the Coulomb gauge operator $\phi_C(\vx)$ transforms as $\phi_C(\vx) \ra e^{\i\th_0}\phi_C(\vx)$, where $\th_0$ is
the $k=0$ mode of the Fourier transformed $\th(k)$. In other words, $\phi_C$ transforms under the remnant global
symmetry in Coulomb gauge.

   The abelian example illustrates an important point with respect to charged states in a gauge theory.  Any physical state
must respect Gauss's Law, which amounts to invariance of the state under infinitesimal gauge transformations.
But one should not conclude from this that all physical states in an infinite volume are entirely gauge invariant, and therefore uncharged.  If that were true then there could be no isolated electric charges in an infinite volume in an abelian theory, even in the massless phase.  Gauss's Law allows a physical state to transform under some global subgroup of the gauge group, and we have seen that a state representing an isolated charge in an abelian theory, i.e.\ $\Psi = \phi_C(\vx) \Psi_0$,
transforms, under an arbitrary gauge transformation, under the remnant global symmetry of the gauge group, providing that
$\Psi_0$ is invariant under that remnant symmetry.  A remnant global
symmetry could be the full group, e.g.\  global U(1) or SU(N), or it could be a subgroup of those
groups.  If a physical state transforms under any of those global symmetries, it is a charged state.  For example, a state in the abelian theory containing $n$ units of electric charge transforms under global U(1)$/Z_n$.  At a minimum, a 
charged state in the non-abelian theory must transform covariantly under the center subgroup of the global gauge group, i.e.\ under $Z_N$ or a non-trivial subgroup of $Z_N$ in the case of SU(N).

\subsubsection{\label{diagonal} The diagonal subgroup}

    If custodial symmetry is unbroken in SU(2) gauge Higgs theory, 
then the full unbroken symmetry is $G \times$SU(2), where $G$ is the remnant global gauge symmetry, and the second factor group is custodial symmetry.  Then if ${\langle \phi \rangle \ne 0}$, which presupposes some gauge choice,  both $G$ and custodial SU(2) are spontaneously broken down to the diagonal subgroup, consisting of transformations
\beq
           \phi'(\vx) = g \phi(\vx) g^\dg ~~,~~ U'_i(\vx) = g U_i(\vx) g^\dg
\label{diag}
\eeq
with $g \in G$.   It is this subgroup of transformations, for ${G=SU(2)}$, which is often referred to in the 
literature as custodial symmetry,  as opposed our terminology (and that of \cite{Georgi:1994qn,Maas:2019nso}), where custodial symmetry refers to the group of global transformations acting on $\phi$ alone.  The diagonal subgroup \rf{diag} 
with $g \in SU(2)$ plays an important role in analyzing the electroweak vector boson mass spectrum, cf.\ \cite{Willenbrock:2004hu,Weinberg:1996kr}.  What is relevant for us is that this unbroken diagonal subgroup, in an $F$-gauge, does not contain the transformation $\phi' = z \phi$, where $z$ is a center element belonging to either the custodial group or the remnant symmetry group, and the vacuum in the broken phase cannot be invariant under field transformations of this kind.

\subsection{The spin glass phase and C confinement}

    In the spin glass phase there are always one or more $F$-gauges in which $\langle \phi \rangle_F \ne 0$.
This means that: (i)  remnant gauge symmetry is broken spontaneously, and in consequence (ii) the vacuum is not a state of definite (zero) color charge; (iii) the color electric field created by a charged field operator is finite; and (iv) the theory
is in a Higgs phase.  Points (i) and (ii) should be obvious, although there may still exist an unbroken diagonal
subgroup of combined remnant gauge and custodial transformations, as mentioned above.
In regard to point (iii), if the energy of the color electric
field created by the field operator in the $F$-gauge were infinite, then that state would be orthogonal to the vacuum, and
$\langle \phi \rangle_F$ would vanish.  Another way to see this is to observe that in an $F$-gauge for 
which $\langle \phi \rangle_F \ne 0$ on every time slice, the $Z_N$ remnant symmetry is broken on every time
slice, and this in turn means that $\langle U_0(\vx,t) \rangle_F$, which transforms under the product $Z_N \times Z_N$
group of remnant symmetry on time slices $t$ and $t+1$, is also non-zero.  One way to think of this is to 
imagine a Monte Carlo simulation in the broken phase of remnant symmetry, where $\phi(\vx,t)$ and $\phi(\vx,t+1)$ fluctuate around fixed backgrounds.  These fixed backgrounds amount to an explicit symmetry breaking background for the $U_0(\vx,t)$ fluctuations, and in consequence $\langle U_0 \rangle_F \ne 0$.  Now define
\beq
           V_F(\vx,\vy,t;U) = g_F^\dg(\vx,t;U) g_F(\vy,t;U)  \ ,
\label{VF}  
\eeq
so that
\bea
   \Psi_V(R) &=& \q^a_F(\vx) q^a_F(\vy) \Psi_0 \non \\
                   &=& \q^a(\vx)  V^{ab}_F(\vx,\vy,t;U) q^b(\vy) \Psi_0 \ .
\eea
Computing $E_V(R)$ in an $F$-gauge, where $V_F=\mathbbm{1}$ and $\langle U_0 \rangle \ne 0$, we find that
\bea
\lim_{R\ra \infty} E_V(R) &=& - \lim_{R\ra \infty}\log \left[ {1\over N} \langle \tr [ U^\dg_0(\vx,t) U_0(\vy,t) ] \rangle_F  \right]  \non \\
              &=& -  \log \left[ {1\over N} \tr [ \langle U^\dg_0(\vx,t) \rangle_F \langle U_0(\vy,t) \rangle_F ] \right] \non \\
              &=& \text{finite} \ ,
\label{EVF}
\eea
which shows that the color electric field carried by charged operators in the $F$-gauge is finite.
The fact that $\langle \phi \rangle_F \ne 0$ means that the vacuum is not an eigenstate of charge (by which we mean,
more precisely, a state which transforms covariantly under the remnant gauge symmetry), and the theory
is in a Higgs phase.  Carrying out the usual expansion with $\phi(x) = \phi_0 + \d \phi(x)$ 
fluctuating around a fixed $\phi_0$, the gauge bosons in U(1) and SU(2) gauge theories all obtain a mass
in the usual way, and there are only Yukawa forces in the theory.  This is C confinement.  

   As a second argument for C confinement, when $\langle \phi \rangle_F \ne 0$, consider the overlap of the charged
and neutral states \footnote{We note that in the SU(2) case the SU(2) group-valued scalar can be  re-expressed at each site as a complex two vector of unit norm, with components $\phi^a(\vx)$ transforming in the fundamental representation \cite{Lang:1981qg}.}
\bea
        |\text{charged}_{\vx \vy} \rangle &=& \q^a(\vx) V_F^{ab}(\vx,\vy;U) q^b(\vy) |\Psi_0 \rangle \non \\
        |\text{neutral}_{\vx \vy} \rangle &=& ( \q^a(\vx)\phi^a(\vx) ) (\phi^{\dg a}(\vy) q^a(\vy) )  |\Psi_0 \rangle \ ,
\label{charged}
\eea
where we imagine taking $\vy \ra \infty$, leaving a quark at site $\vx$.
These are both physical states, but the neutral state is obtained from operators creating two separated color singlet objects, with
no color electric field diverging from points $\vx,\vy$.  The charged state is created by operators which transform, at sites
$\vx,\vy$, under the remnant gauge symmetry.  Then evaluating the overlap in the
$F$-gauge, where $V_F=\mathbbm{1}$, integrating out the heavy quark fields, and taking the $R=|\vx-\vy| \ra \infty$ limit
\bea
       \lim_{|\vx-\vy| \ra \infty} \langle \text{neutral} | \text{charged} \rangle 
      &\propto&    \lim_{|\vx-\vy| \ra \infty} \langle  \phi^{\dg a}(\vx) \phi^a(\vy) \rangle_F \non \\
     &=& \langle \phi^{\dg a} \rangle_F \langle \phi^a \rangle_F \non \\
      &>& 0 \ .
\eea
This non-zero overlap shows that the ``charged'' state containing an isolated quark at point $\vx$ is not really charged; it has a finite overlap with states created by color singlet operators acting on the vacuum at point $\vx$.  If the state created by a color singlet operator is neutral, then so is the state created by the charged operator.   In fact, we see that the ``charged'' state
is not associated with a long range color electric field characteristic of a charged field.  If it were, and because there is no such
long range field in the state created by a color singlet operator acting on the vacuum, then the charged and neutral states would be orthogonal, which is not the case.

    However, in the spin glass phase there also exist $F$-gauges in which $\langle \phi \rangle_F = 0$ in some or all of this region of the phase diagram.  But it makes no
sense to say that the theory is in a Higgs phase in one $F$-gauge but not in another.  Either the system is in a C confinement
phase or it is not; this is a question which is independent of the gauge choice.  The vanishing of the Higgs field expectation
value $\langle \phi \rangle_F$ does imply that the state created by $\phi_F$ operating on the vacuum is orthogonal to the vacuum, but this could be for one of two reasons:
(a) the vacuum is an eigenstate of charge, and remnant gauge symmetry is unbroken; or (b) the
color electric field created by $g_F(\vx;U)$ is of infinite energy, and for this reason the overlap of the 
state $\phi_F \Psi_0$ with $\Psi_0$ can vanish.
Option (a) seems inconsistent in the spin glass phase, for reasons just mentioned, and also because the remnant $Z_N$ gauge symmetry is indistinguishable, in its action on the $U,\phi$ fields, from the $Z_N$ subgroup of custodial symmetry, and the order parameter for broken custodial symmetry is gauge invariant.   Consistency therefore requires that if $\langle \phi \rangle_F=0$ in some region of the spin glass phase in one particular $F$-gauge, then also $\langle U_0 \rangle_F = 0$ in that gauge in the same region.  In that case, the last line of \rf{EVF} should be infinity, and option (b) is the correct explanation.    

\subsubsection{Example: Axial gauge}

     An example of $\langle \phi \rangle_F=0$ in the spin glass phase is the maximally fixed axial gauge  
$U_1(\vx,t)=\mathbbm{1}$, with $U_2, U_3$ set to $\mathbbm{1}$ on a plane and a line, respectively, to eliminate any residual local gauge symmetry on a timeslice.  We can show that both of the expectation values of $\phi$ and $U_0$ vanish in this gauge, and that the charged field operators $q,\q,\phi$ create infinite energy states.  Let the subscript $A$ denote the axial gauge just mentioned.  Then the transformation
to the gauge is
\beq
          g_A(\vx,t;U) = \left\{ P\left[\prod_{n=0}^\infty U_1(\vx+n\ihat,t) \right] \right\}^\dg \ ,
\eeq
where $P$ denotes path ordering in the $x$-direction. Defining $V_A$ as in \rf{VF}, and supposing that $\vx$ and 
$\vy=\vx+R\ihat$ lie on a line parallel to the $x$-axis, then
$V_A(\vx,\vy,U)$ is simply a Wilson line joining points $\vx,\vy$, i.e.
\beq
          V_A(\vx,\vy,t;U) = P\left[\prod_{n=0}^{R-1} U_1(\vx+n\ihat,t) \right] \ .
\eeq
Therefore
\beq
          E_V(R) = -\log\left[{1\over N}W(R,1)\right] \ ,
\eeq
where $W(R,1)$ is the expectation value of a rectangular timelike Wilson loop which is one lattice spacing long
in the time direction.  Since $W(R,1)$ falls exponentially to zero with the perimeter $2R+2$, it follows that
\beq
          \lim_{R\ra \infty} E_V(R) = \infty  \ .
\label{WR}
\eeq
On the other hand, if $E_V(R)$ is evaluated in axial gauge, where $V(\vx,\vy,U) = \mathbbm{1}$, then for $\vx-\vy$ parallel
to the $x$-axis we have in this gauge
\bea
        \lim_{R\ra \infty} E_V(R) &=& - \lim_{R\ra \infty} \log\left[{1\over N}
                           \langle \tr U_0^\dg(\vx,t) U_0(\vy,t) \rangle_A \right]  \non \\
              &=& -  \log\left[{1\over N} \tr \langle U_0^\dg(\vx,t)\rangle_A \langle U_0(\vy,t) \rangle_A \right]  \ .
\eea
From \rf{WR} we see that, in axial gauge, $\langle U_0^\dg(\vx,t)\rangle_A = 0$.  It follows that the state created by
$q,\q$ operators in axial gauge, i.e.
\bea
           \Psi_{V}(R) &=& \q(\vx) V_A(\vx,\vy;U) q(\vy) \Psi_0 \non \\
                              &=&   \q_A(\vx) q_A(\vy)  \Psi_0 \ ,
\eea
is infinite energy in the $R \ra \infty$ limit.  Moreover, from ${\langle U_0\rangle_A = 0}$ we deduce that
$\langle \phi \rangle_A = 0$, because if this were not so, then $U_0$ would have a finite expectation value.  We conclude that
$\langle \phi \rangle_A = \langle U_0\rangle_A = 0$ in axial gauge, and isolated field operators $\phi,q,\q$ create infinite energy states in this gauge.  For axial gauge this is actually true at all couplings.  The underlying reason is that a Wilson line operator creates a line of electric flux whose energy increases with length, regardless of whether the theory is in a C, \Sc \, or massless phase.   

\subsection{The symmetric phase and \Sc \ confinement}   
     
      If  custodial symmetry is unbroken in the ground state, then for any operator $Q(\phi,U)$ composed of fields on a time slice, and $z$ in the center subgroup of custodial symmetry,
\beq
           \langle Q(z \phi,U) \rangle_F   =   \langle Q(\phi,U) \rangle_F
\eeq
in the appropriate $h\ra  0$ and infinite volume limits.  But the action of the $Z_N$ subgroup of remnant gauge symmetry on fields $\phi,U$ is indistinguishable from the action of the $Z_N$ subgroup of custodial symmetry, so if custodial symmetry is unbroken, so is
the $Z_N$ remnant gauge symmetry, and the vacuum state is invariant under this symmetry.  This means that any operator which transforms covariantly under the $Z_N$ remnant gauge symmetry, operating on the vacuum, creates a charged state which also transforms covariantly under $Z_N$ remnant gauge symmetry.   States of this kind can be created, e.g., by
operators $q_F,\q_F,\phi_F$ in any $F$-gauge, which transform covariantly under (at a minimum) the remnant $Z_N$ symmetry.   
  
     With this in mind we return to the overlap of charged and neutral states, as defined in \rf{charged}, this time in the
symmetric phase.  We have
\bea
     & &  \lim_{|\vx-\vy| \ra \infty} \langle \text{neutral} | \text{charged} \rangle \non \\
      & & \qquad \propto    \lim_{|\vx-\vy| \ra \infty} \langle  \phi^{\dg a}(\vx) V^{ab}(\vx,\vy;U) \phi^b(\vy) \rangle \non \\
      & & \qquad  = \lim_{|\vx-\vy| \ra \infty} \int DU \overline{\phi^a(\vx) \phi^b(\vy)}[U] V^{ab}(\vx,\vy;U) P(U) \ , \non \\
\eea
where
\bea
 \overline{\phi^a(\vx) \phi^b(\vy)}[U]  &=& {1\over Z_{spin}(U)} \int d\phi \phi^a(\vx) \phi^b(\vy) e^{-H_{spin}/kT} \ . \non \\
\eea
Since custodial symmetry is unbroken for gauge configurations drawn from the probability distribution $P(U)$, it follows that
for such configurations, in the symmetric phase at $h \ra 0$,
\beq
       \lim_{|\vx-\vy| \ra \infty} \overline{\phi^a(\vx) \phi^b(\vy)}[U] = 0 \ .
\eeq
Because $V$ is a bounded operator (see \rf{norm}), this means that the overlap between all charged states and the
neutral, ``string broken'' states must vanish in the $R\ra \infty$ limit:
\beq
 \lim_{|\vx-\vy| \ra \infty} \langle \text{neutral} | \text{charged} \rangle = 0 \ .
\eeq
Note that this result holds for all $V$ operators in the symmetric phase, independent of any gauge choice.
  
     Charged states in the unbroken phase may be of either finite or of infinite energy. For example, if there is an $F$-gauge in the unbroken phase such that $\langle U_0 \rangle_F \ne 0$, then the energy 
$E_{V_F}$ at $R\ra \infty$ is finite, according to \rf{EVF}.\footnote{This is not inconsistent with $\langle \phi \rangle_F = 0$.  While $\langle \phi \rangle_F \ne 0$ implies $\langle U_0 \rangle_F \ne 0$, the converse is not necessarily true.} If there are no charged states of finite energy above the vacuum energy, then the system is in an \Sc \ confinement phase.  If, on the other hand, there do exist charged finite energy states, orthogonal to all neutral states, then states of this kind will necessarily appear in the spectrum.  The system cannot then be in a C confining phase, where there are no charged particles in the spectrum.  Nor can it be in an \Sc \ confining phase, where isolated charges
are all states of infinite energy.  The remaining possibility is a massless phase.  So the phase of unbroken custodial symmetry
is either \Sc \ confining, or massless.  This is consistent with the fact that $\langle \phi \rangle_F=0$ in all $F$-gauges in the symmetric phase, so there exists no sensible perturbative expansion
of $\phi(x)$ around a non-zero expectation value, and no broken symmetry that could supply a $1/k^2$ pole in the scalar propagator, which could then be absorbed to produce a massive pole in gauge boson propagators.  In other words, there is no Brout-Englert-Higgs mechanism in the symmetric phase, at least not one that can be seen in any physical $F$-gauge.

\subsubsection{Pseudo-matter fields}

    Charged states may also be created, in the unbroken phase, by combining matter fields with ``pseudo-matter'' fields.
A pseudo-matter field (cf.\ \cite{Greensite:2017ajx}) is an operator $\omega^a(\vx;U)$ which is entirely a functional of the gauge field, transforming in the fundamental representation of the gauge group for all local
gauge transformations, but which is invariant under global $Z_N$ transformations.  An explicit example, in the continuum abelian gauge theory, is $g_C(\vx;A)$ in \rf{Dirac},
which transforms covariantly under local gauge transformations, but is invariant under global U(1) gauge transformations.   
Another example is any eigenstate $\zeta^a_n(\vx;U)$, or any linear combination of eigenstates, of the spatial covariant Laplacian
\beq
           (-D_i D_i)^{ab}_{\vx \vy} \zeta^b_n(\vy;U) = \l_n \zeta^a_n(\vx;U)  \ ,
\eeq
where
\bea
 \lefteqn{(-D_i D_i)^{ab}_{\vx \vy} = } \non \\
    &=& \sum_{k=1}^3 \left[2 \d^{ab} \d_{\vx \vy} - U_k^{ab}(\vx) \d_{\vy,\vx+\hat{k}}    - U_k^{\dg ab}(\vx-\hat{k}) \d_{\vy,\vx-\hat{k}}   \right] \ . \non \\  
\eea
Since the gauge field is invariant under global $Z_N$ gauge transformations, the same is true of the 
$\zeta_n(\vx;U)$ eigenstates, although these operators must transform covariantly in the fundamental representation under all other gauge transformations.   Given a pseudo-matter field $\omega^a(\vx;U)$, a charged state in the unbroken phase, associated with the operator $V(\vx,\vy;U)$ in the $\vy \ra \infty$ limit, can be constructed as in \rf{PsiV} by replacing the quark operator $q^a(\vy)$ with a pseudo-matter operator, creating a state
\bea
         \Psi_{q} &=& \q^a(\vx) V^{ab}(\vx,\vy;U) \omega^b(\vy;U) \Psi_0 \non \\
                      &=&  \q^a(\vx)   \widetilde{\omega}^a(\vx;U) \Psi_0 \ ,
\label{Psiq}
\eea
and taking the same $\vy \ra \infty$ limit.   Note that
\beq
  \widetilde{\omega}^a(\vx;U) =  V^{ab}(\vx,\vy;U) \omega^b(\vy;U) 
\eeq
is itself a pseudo-matter operator, and indeed any state of the form \rf{Psiq}, for any pseudomatter field  
$\widetilde{\omega}^a(\vx;U)$,  transforms under global $Z_N$ gauge transformations, but not under local gauge transformations.

    A set of $N$ pseudo-matter fields $\omega^a_n(\vx;U)$   can be used to define an
$F$-gauge choice, as exemplified by the Laplacian gauge introduced by Vink and Wiese \cite{Vink:1992ys}. 
Following those authors, define
\bea
           M^{ab}(\vx) &=& \omega^a_b(\vx;U) \ ,
\eea
and carry out a polar decomposition at each site
\beq
           M(\vx;U) = W(\vx;U) P(\vx;U) \ ,
\eeq
where $W$ is a unitary matrix.  Let 
\beq
D(\vx;U) = e^{i\a(\vx)/N} \mathbbm{1} ~~,~~ e^{i\a(\vx)} = \det[W(\vx;U)] \ .
\eeq
Then the SU(N) matrix-valued field
\beq
            g_F(\vx;U) = D(\vx;U) W^\dg(\vx;U)
\eeq
defines the gauge transformation to an $F$-gauge. In this construction the remnant gauge symmetry is reduced to the minimal symmetry possible, i.e.\ to the center subgroup $G=Z_N$, and
the gauge condition on any time slice is
\beq  
      F(\vx,U) = g_F(\vx;U) - \mathbbm{1} = 0 \ .
\eeq
The Gribov ambiguity is eliminated in gauges of this kind, and the components of, e.g., the $\q$
operator in this gauge are given by
\beq
            \q_F^a(\vx) = \q^b(\vx) g_F^{ba}(\vx;U)  \ .
\eeq
Each of the $\q_F^a(\vx)$ components is invariant under local gauge transformations, and transforms under the
global remnant $Z_N$ symmetry.  This remnant symmetry does not transform the components among themselves.  

   In refs.\ \cite{Frohlich:1981yi,tHooft:1979yoe} it was observed that particles in the physical spectrum of an SU(2) gauge
Higgs theory (such as the physical $W$'s, quarks, and Higgs particles)  are created from local gauge invariant composite operators, and these
particles are all color singlets.  That observation is correct if there is no massless phase in the theory, and the charged operators
create only infinite energy states in the unbroken phase.  If there is however a massless
phase, as in the abelian Higgs model in 3+1 dimensions and lattice SU(2) gauge Higgs theory in 4+1 dimensions, then there are charged particles in the spectrum in that phase, 
and the list of operators in \cite{Frohlich:1981yi,tHooft:1979yoe} is incomplete.  As
we have just pointed out, it is possible to construct physical charged states, invariant under local but transforming under global
gauge transformations, and these are required to complete the spectrum in the massless phase.
 
\subsection{Transitions}
     
    To summarize:   Any physical $F$-gauge, in which field operators $\phi,q,\q$ acting on the vacuum create physical states,
leaves unfixed a remnant gauge symmetry containing at least the global $Z_N$ subgroup. In the spin glass phase this global $Z_N$ subgroup is broken spontaneously, and the \Sc \ confinement condition is violated.  Physical states in this phase cannot be distinguished by their transformation properties under the remnant gauge symmetry; there are no charged states in the spectrum distinct from neutral states. This is the Higgs, or C confinement phase.  In this phase there always exist $F$-gauges in which the field operators create finite energy states, and $\langle \phi \rangle_F$ is non-zero. 

In the phase of unbroken custodial symmetry the global $Z_N$ subgroup of gauge symmetry is unbroken, and 
$\langle \phi \rangle_F$ vanishes in every physical gauge.  In this phase it is possible to construct charged states orthogonal to the vacuum, and orthogonal to any uncharged state, which transform covariantly under the remnant gauge symmetry.  If there exist charged states of finite energy, then charged states must appear in the spectrum, in which case the theory is not in a C confinement phase, and \Sc \ confinement is also ruled out. The remaining possibility is a massless phase. But if every charged state is a state of infinite energy relative to the vacuum,  then there is separation-of-charge confinement, and the system is in the \Sc \ confined phase.         
   
    The conclusion is that the spin glass phase is a C confinement Higgs phase, while the phase of unbroken custodial
symmetry may be either a massless or an \Sc \ confining phase, depending on the couplings.   In the absence of a massless phase, as in SU(2) gauge Higgs theory in $D=3+1$ dimensions, {\it the transition to the spin glass phase coincides with the transition from \Sc \ confinement to C confinement.} \\

    That is the main result of  this paper.

 \subsection{\label{other} Other gauges}
 
    In unitary gauge, in U(1) and SU(2) gauge Higgs theories with a fixed modulus Higgs field, we set can set 
$\phi(x) = \mathbbm{1}$ everywhere.  Then a standard perturbative analysis suggests that the theory is in a Higgs
phase everywhere in the phase diagram, at all $\b, \g$ couplings apart from $\g=0$.  This conclusion is demonstrably false in some regions of the phase diagram, since it is known from numerical simulations that the abelian Higgs model with
a compact gauge group has a massless phase in $D=3+1$ dimensions \cite{Ranft:1982hf,Matsuyama:2019lei},
while the SU(2) gauge Higgs model has, on the lattice, 
a massless phase in $D=4+1$ dimensions \cite{Beard:1997ic}.  The question
is what is going wrong with the usual perturbative reasoning in unitary gauge.

    The answer is that field operators transformed to unitary gauge (denoted by a subscript $U$) are color singlet operators, and the non-zero expectation value of the scalar field operator $\langle \phi \rangle_U$, which is a triviality in unitary gauge, says nothing about the vacuum state and the phase of the theory.  To see this, we observe how the Gauss law operates in unitary
gauge.  In this case the transformation $g_U$ to unitary gauge, in U(1) and SU(2) gauge theories with unimodular Higgs fields, is simply
\beq
            g_U(\vx;\phi) = \phi^\dg(\vx)
\eeq
so that 
\beq
          \phi_U(\vx) = g_U(\vx;\phi) \phi(\vx) = \mathbbm{1}
\eeq
is an uncharged, color singlet operator.  The same observation holds for other matter field operators in unitary gauge,
\bea
           q_U^a(\vx) &=& \phi_U^{\dg ab}(\vx) q^b(\vx) \non \\
           \q_U^a(\vx) &=&  \q^b(\vx) \phi^{ba}_U(\vx) \ ,
\eea
which are also local color singlets, invariant under all gauge transformations including transformations in the global 
center subgroup.  Isolated color neutral operators of this type, operating on the vacuum state, will produce excited states of finite energy at any $\b,\g>0$ in phase plane, but even in the massless phase they cannot by themselves create charged states.

     Another question which arises in unitary gauge is the fate of the dynamical degrees of freedom associated with custodial
symmetry, which seem to have disappeared in this gauge.  That disappearance is deceptive, however.  In fact the relevant degrees of freedom are still there in unitary gauge, but they are now found in the gauge sector.  Let us write 
$U_\m(x) = g(x) U_{\m,F}(x) g^\dg(x+\hat{\m})$, where $U_{\m,F}$ is the gauge field in an $F$-gauge, and $g(x)$ is some
SU(2) valued field.  Custodial symmetry is now a global transformation on the $g(x)$ field.  To see this, begin by fixing to unitary gauge, $\phi=\mathbbm{1}$.  Then, in the SU(2) gauge Higgs model \rf{Sgh},
\beq
          Z = \int DU \exp\Bigl[-S_W + \g \sum_{x,\m} \oh \tr [U_\m(x)]\Bigr] \ .
\label{Sunit}
\eeq
Now insert unity in the usual way to obtain
\bea
         Z &=& \int DU \left\{ \D_{FP}[U] \int Dg \d(F[g \circ U]) \right\} \non \\
             & & \, \times \exp\Bigl[-S_W + \g \sum_{x,\m} \oh \tr U_\m(x)\Bigr]  \non \\
            &=& \int DU  \D_{FP}[U]  \d(F[U]) e^{-S_W}  \non \\
            & & \, \times \int Dg \exp\Bigl[ \g \sum_{x,\m} \oh \tr [g^\dg(x) U_\m(x) g(x+\hat{\m})\Bigr] \ ,
\eea      
which is simply the SU(2) gauge Higgs theory in an $F$-gauge.   Obviously the Higgs action $S_H$ is again 
invariant under custodial transformations of the $g$ field, and unbroken custodial symmetry, combined with 
$E_V(R)$ finite for some $V$ at $R\ra \infty$, is a sufficient condition for the existence of a massless phase, 
as discussed above.  Similar considerations apply to the abelian Higgs model.

The action in unitary gauge in \rf{Sunit}, for the U(1) and SU(2) gauge groups, is sometimes described as a rewriting of the action in terms of gauge invariant variables.  While it is true that the link variables in unitary gauge are formally gauge invariant, it can be mistaken to conclude, simply from the form of the action in \rf{Sunit}, that the theory is necessarily in a massive Higgs phase, regardless of the couplings $\b,\g$. This conclusion, as already mentioned, is directly contradicted by numerical simulations.

    Turning to covariant gauges such as Landau gauge, the problem is that the gauge condition cannot be imposed
independently on each time slice, and the transformation to e.g.\ Landau gauge, $g_{Lan}(\vx,t,U)$,
is a function of the gauge field over the entire lattice, on all time slices. The construction of a $\Psi_V$ state with
$V=V_F$ in \rf{VF} does not work in any covariant gauge, and the energetics argument in eq.\ \rf{EVF} does not apply.
 The same true in temporal gauge, where the  transformation
$g_{temp}(x,U)$ to temporal gauge on the Euclidean lattice  involves the $U_0$ component of the gauge field, and physical states are not produced by isolated field operators $\phi,q,\q$  acting on the vacuum.  Unless
these field operators are combined with pseudo-matter fields, as described above, such operators will generate unphysical states that violate the Gauss law constraint.  In covariant gauges, an isolated field operator acting on the vacuum will also produce
an unphysical state, but in fact even the physical state conditions in covariant gauges are not well defined on the lattice,
for reasons associated with Neuberger's theorem \cite{Neuberger:1986xz}.  The theorem tells us that the expectation values
of BRST invariant operators are formally ratios of $0/0$.  This ambiguity is evaded in lattice Monte Carlo simulations, as already noted in section \ref{gauge}, by a restriction, implemented by the gauge fixing algorithm, of gauge field configurations to the first Gribov region.  But this restriction breaks BRST symmetry \cite{Cucchieri:2014via}, so the identification of physical states in the lattice formulation is  problematic in covariant gauges.     

   For all of these reasons we have focussed on the implications of $\langle \phi \rangle$ in the physical $F$-gauges, since the
unitary, temporal, and covariant gauges do not seem suitable for our purposes.

\section{Discussion}

\subsection{Comparison to our earlier proposal}

   The custodial symmetry breaking criterion put forward in section \ref{spin} is a little different from our original proposal
in \cite{Greensite:2018mhh}, so we should explain the difference.  The two criteria are very similar, but here we define the spin glass phase in $D$ space dimensions in a quantum statistical system defined by the quantum Hamiltonian $H$ and 
the operator $\exp[-H/kT]$, with an order parameter $\Phi(U)$ that depends only on link variables on a time slice, whereas in our previous work we defined the spin glass phase in $D+1$ Euclidean spacetime dimensions, and a classical Boltzmann weight $\exp[-S_E]$, where the order parameter depends on the gauge fields at all times.  The difference in the Monte Carlo implementations is this:  in the current proposal, when computing the spin glass order parameter by means of the
$D+1$ dimensional Euclidean path integral, the gauge field is fixed only on the $t=0$ time slice, but is allowed to vary on all other time slices in the Euclidean path integral, while in our previous formulation 
the spin glass order parameter\footnote{Described as such in \cite{Greensite:2018zeo} in the context of an abelian lattice gauge theory.} was computed in the same Euclidean path integral, but the gauge field was fixed at all times. Likewise, in our earlier work, the order parameter was computed from $\phi(\vx,t)$ at all times, while in the present case the order parameter
is computed only from  $\phi(\vx,t)$ on the $t=0$ time slice. We conjectured, in our previous work, that custodial symmetry breaking coincides with the \Sc \ to C confinement transition.  That conjecture was almost correct.  The statement is true in our new formulation, as set out in this paper. 

\subsection{Goldstone's Theorem}

    Custodial symmetry is a continuous global symmetry, which raises the question of why there are no gapless
excitations in the spin glass phase.  The answers are similar to those provided by Guralnik et al.\  
\cite{Guralnik:1967zz} many decades ago, as we have already remarked in section \ref{gauge}.  
In the first place, any long range correlations in $\phi(\vx)$ at fixed $U_i(\vx)$ are gauge dependent, and, in the absence of gauge-fixing, average to zero after integrating over $U_i(\vx)$.   In the second place, upon fixing to a physical $F$-gauge, the Hamiltonian operator is in general non-local, which violates one of the assumptions that goes into the proof of the Goldstone theorem.  

\subsection{$Z_N$ symmetry}

   We note again the relevance  of the non-trivial center of the gauge group.  If the center is trivial, 
then there are local operators  $\xi^a(\vx;U)$, transforming in the fundamental representation of the gauge group, which depend only on the gauge field.   Unlike the pseudo-matter operators described above, these operators transform under all elements
of the gauge group. Then, choosing ${V^{ab}(\vx,\vy;U) = \xi^a(\vx;U) \xi^{\dg b}(\vy;U)}$,  the corresponding $\Psi_V$ state
consists of two separated color singlet excitations above the ground state, whose interaction energy is negligible,
and \Sc \ confinement is ruled out.  

   Suppose instead that the center subgroup is non-trivial, but that the scalar field transforms in a zero $N$-ality representation,
such as the adjoint representation of the gauge group, as in the Georgi-Glashow model.   A custodial symmetry, if one exists,
does not necessarily contain a subgroup which coincides with a global subgroup of the gauge transformations.  The \Sc \
criterion still makes sense, for $q,\q$ static quarks transforming in the fundamental representation, but in this case it is associated with a different group of $Z_N$ transformations which are not gauge transformations, and not elements of the custodial group.  This is the center symmetry whose importance to confinement in pure gauge theories was emphasized long ago by 't Hooft \cite{tHooft:1977nqb}, and which is associated with the center vortex theory of 
confinement (cf.\ the review in \cite{Greensite:2011zz}).  The order parameter for the breaking of this symmetry is the Polyakov line.   The system is in an \Sc \ confinement phase if and only if this global $Z_N$ center symmetry
is not spontaneously broken.

\subsection{Custodial symmetry and the spectrum}

    Particles in the physical spectrum of a gauge Higgs theory in the C and \Sc \ phases are created by local color singlet operators, as pointed out long ago in  \cite{Frohlich:1981yi,tHooft:1979yoe}.  In SU(2) gauge Higgs theory there is a triplet  of massive vector bosons, the $W$ bosons, that can be created by such operators in the Higgs phase.  This is in accordance with the usual perturbative treatment, where there is a massive vector boson associated with each of the three ``broken'' generators of the gauge group.
However, as emphasized recently by Maas and T\"{o}rek \cite{Maas:2018xxu}, this agreement between the perturbative counting of asymptotic particle states, and the distinct states that can be created by local gauge invariant operators, seems to be a coincidence in SU(2) gauge Higgs theory, resulting from the fact that the custodial group is also SU(2).  For larger gauge groups the counting of vector boson states according to the perturbative BRST approach, and the number of distinct vector bosons that can be created by local color singlet operators according to the analysis of  \cite{Frohlich:1981yi,tHooft:1979yoe}, may not agree,  cf.\ \cite{Maas:2018xxu}.

\subsection{Other Proposals}

    Our spin glass criterion is certainly not the first proposal for distinguishing the Higgs phase from other phases,
so it may be appropriate to comment here on other suggestions, many of which are found in the
condensed matter literature.

    Most modern textbooks on quantum field theory and many-body theory have a section on spontaneously broken
gauge symmetry, and in most cases what is done is to fix to unitary gauge and show that the would-be Goldstone
mode is ``eaten'' by the gauge particles, which then acquire a mass.   We have already discussed, in section \ref{other}, the deficiencies in this argument.  In fact, the subject heading ``spontaneous gauge symmetry breaking'' is already a little misleading, since no {\it local} gauge symmetry can break spontaneously, as we know from the Elitzur theorem.

    However, the Elitzur theorem does not forbid the spontaneous breaking of a global remnant of the gauge symmetry in some definite gauge, and therefore some treatments, e.g.\ \cite{Duncan}, define the concept of spontaneous gauge symmetry breaking as the breaking of that remnant symmetry, deduced from $\langle \phi \rangle \ne 0$ in some gauge.  The problem is that the transition to the Higgs phase would then appear
to depend on the gauge choice, e.g.\ the transition lines in Coulomb and Landau gauge do not coincide \cite{Caudy:2007sf}. Physics can't depend on gauge choice, so the absence of a Higgs phase cannot be inferred simply from 
$\langle \phi \rangle = 0$ in one particular gauge.   

    In the lattice gauge theory literature, the Osterwalder-Seiler-Fradkin-Shenker work
\cite{Osterwalder:1977pc,Fradkin:1978dv}, combined with the observation that physical particles in the Higgs sector are
created by local gauge invariant operators \cite{Frohlich:1981yi,tHooft:1979yoe}, have generally discouraged any attempt to distinguish the Higgs from the confinement region, when the scalar field is in the fundamental representation of the gauge group.  The Fredenhagen-Marcu confinement criterion \cite{Fredenhagen:1985ft}, for example, was intended as a definition of confinement in a theory with matter in the fundamental representation, but this criterion
essentially amounts to distinguishing massive from massless phases.  It does not distinguish the Higgs from the
confinement phase, which both satisfy the Fredenhagen-Marcu criterion.  In other words, it is essentially a criterion
for C confinement.

    In many-body theory it becomes more urgent to distinguish between the normal (i.e.\ massless) phase,
and the superconducting (``Higgs'') phase.  Of course the superconducting phase in the abelian theory differs from the normal phase by, e.g., the Meissner effect, and also by certain topological properties \cite{Hansson:2004wca}.   However, we are interested in whether these effects are associated with a broken continuous symmetry.
In a simple BCS Hamiltonian which ignores any coupling to the gauge field, or the corresponding Landau-Ginzburg model derived from a Hubbard-Stratonovich transformation, there is only a global U(1) symmetry which, if spontaneously broken, results in finite expectation values for the Cooper pair creation operator 
$c^\dg_\ua(x) c^\dg_\da(x)$ or the double-charged scalar field $\phi(x)$.  The problem is that when the electrons, or the scalar field, are coupled to a quantized electromagnetic field, and the theory becomes locally gauge invariant, an expectation
value for these charged operators is ruled out by the Elitzur theorem.  To deal with this difficulty, there are a number of proposals which would replace the gauge non-invariant order parameter by an ostensibly gauge invariant order parameter.  When these order parameters are examined closely, they always boil down to a gauge choice, in the sense that if these
order parameters are evaluated in a particular gauge, they reduce to $\langle \phi \rangle$, or (what amounts to the same thing) the correlator $\langle \phi^\dg(x) \phi(y) \rangle$ in the $|x-y| \ra \infty$ limit.   An example in ordinary scalar QED, where the charge of the scalar $\phi$ field is an integer 
multiple of electric charge $ne$, is the operator construction due to Dirac,
\bea
           \Omega(\vx) &=&  g_C^n(\vx;A) \phi(\vx) \ ,
\label{Dorder}
\eea
where $g_C$  is the gauge transformation to Coulomb gauge, shown explicitly in \rf{Dirac}.  This order parameter
is invariant under local gauge transformations, but not under global gauge transformations $g(\vx) = e^{i\th}$, which is the remnant gauge symmetry in Coulomb gauge, as already noted.   In Coulomb gauge, $\Omega(\vx)=\phi(\vx)$.  Other proposals for locally gauge invariant order parameters, constructed along the same lines, are based on Lorentz gauge \cite{Kennedy:1986ut}, or (implicitly) on axial gauges \cite{schakel,dutch}.  The point is that all of these order parameters depend at least implicitly on a gauge choice
for their construction.\footnote{It has also been suggested in \cite{GREITER2005217} that the symmetry which is broken in the abelian theory is a global U(1) symmetry which is distinct  from global gauge symmetry.  There may very well be a connection here to custodial symmetry breaking, although in the absence of an order parameter which would detect the breaking of this distinct symmetry we are not able to make an appraisal.}  Where there is a thermodynamic transition, these parameters sometimes (but not always \cite{Matsuyama:2019lei})  agree on the location of the transition.  In regions of the phase diagram where there is no thermodynamic transition, in both abelian and non-abelian lattice gauge Higgs theories, such order parameters will still locate transition lines, but will in general disagree on their locations \cite{Caudy:2007sf,Matsuyama:2019lei}.   The point is that if $\langle \phi \rangle_F \ne 0$ in a physical $F$-gauge, then the theory is in a Higgs phase.  
But if $\langle \phi \rangle_F = 0$ the theory may or may not be in a Higgs phase, as discussed at length above.

\section{Conclusions}

   We have shown that gauge Higgs theories possess a spin glass phase in which a global custodial symmetry is broken
spontaneously, and that the transition to the spin glass phase in a non-abelian theory, in the absence of a massless phase, is accompanied by a transition from one type of confinement to another, namely  a transition from separation-of-charge (\Sc)  confinement to ordinary color (C) confinement.   The asymptotic  particle spectrum in both phases consists of color singlets, but this is because of broken symmetry in the C confinement phase, and for energetic reasons in the \Sc \ phase. The \Sc \ phase can only exist for gauge groups with a non-trival center subgroup.

It therefore seems meaningful to identify the Higgs phase of gauge Higgs theory as the spin glass phase, in which a global custodial symmetry is spontaneously broken, and which is distinguished by both symmetry and confinement type from a non-spin glass or confinement phase.  These qualitative distinctions between the Higgs and confinement phases exist even in the absence of a thermodynamic transition which completely isolates these phases from one another, and the symmetry breaking order parameter does not involve, either explicitly or implicitly, a choice of gauge.

   Since the color electric field energy associated with a pair of charged objects grows with separation in the \Sc \ confinement phase, up until string breaking by matter fields, a spectrum of resonances associated with color flux tube formation, lying on linear Regge trajectories, seems inevitable.  The mechanism is the same as in QCD: there is some energetic barrier to pair production and string breaking, and even when energetically favorable, string breaking is not immediate.  Hence the existence of flux tube resonances in the \Sc \ phase.  In the spin glass phase, where one can always find a physical gauge in which $\langle \phi \rangle_F \ne 0$, particle pair production is not really required for string breaking, and this energetics argument may not apply.  It is in any case an experimental fact that there is no spectrum of resonances of this kind in the electroweak theory.

    The transition between the \Sc \ confinement and Higgs phases also represents the boundary between a region where
a perturbative approach may apply, and a region where such an approach must fail.  The growth of energy with color charge separation, which is the definition of \Sc \ confinement, is fundamentally non-perturbative, as it is in QCD.  Moreover, there is no physical $F$-gauge, in the \Sc \ confinement phase, for which $\langle \phi \rangle_F \ne 0$, so the expansion of the Higgs field around some non-zero minimum in the Higgs potential is almost certainly misleading.  This expansion can only make sense in the Higgs region, at least in physical gauges.  It must fail in the \Sc \ confining and (in the lattice abelian Higgs model) the massless phases.

   Using the procedures described here, it should be possible to map out numerically the confinement/Higgs phase structure for SU(N) gauge Higgs systems with one or more scalar fields, and this may conceivably have phenomenological implications.  We hope to return to this question at a later time. \\

\acknowledgments{J.G.\ would like to thank Maarten Golterman, Kristan Jensen, and Axel Maas for helpful discussions. This research is supported by the U.S.\ Department of Energy under Grant No.\ DE-SC0013682.}   

\bibliography{sym3}

\end{document}